\documentclass[aps,secnumarabic,nofootinbib,amsmath,amssymb,superscriptaddress,nobibnotes,floatfix,twocolumn,showpacs]{revtex4-1}
\usepackage[table]{xcolor}
\usepackage{lipsum}
\usepackage{float}
\usepackage[normalem]{ulem}
\usepackage{cancel} 
\usepackage{verbatim}
\usepackage{esint}
\usepackage{mathrsfs}
\usepackage{comment}
\usepackage{braket}
\usepackage{bbold}
\usepackage{xcolor}
\usepackage{color}
\usepackage[latin1]{inputenc}
\usepackage{graphicx}
\usepackage{dsfont}
\usepackage{amsmath,amsfonts,amssymb,amsthm}
\usepackage{xspace}
\usepackage{times}
\usepackage{longtable}
\usepackage{extarrows}
\usepackage{multirow}
\usepackage{threeparttable}
\usepackage{ulem}   
\usepackage[colorlinks, linkcolor=blue, anchorcolor=blue, citecolor=blue]{hyperref}

\newcommand\AAA{\textit{A}}
\newcommand\BBB{\textit{B}}
\newcommand\ccc{\textit{c}}
\newcommand\ddd{\textrm{d}}
\newcommand\EEE{\textit{E}}
\newcommand\eee{\textit{e}}
\newcommand\FFF{\textit{F}}
\newcommand\fff{\textit{f}}
\newcommand\kkk{\textit{k}}
\newcommand\mmm{\textit{m}}
\newcommand\NNN{\textit{N}}
\newcommand\nnn{\textit{n}}
\newcommand\ppp{\textit{p}}
\newcommand\qqq{\textit{q}}
\newcommand\RRR{\textit{R}}
\newcommand\rrr{\textit{r}}

\newcommand\ttt{\textit{t}}
\newcommand\uuu{\textit{u}}
\newcommand\VVV{\textit{V}}

\newcommand\WWW{\textit{W}}

\newcommand\PPP{\textit{P}}

\begin{document}


\title{Acceleration of ultra-high energy cosmic rays in the early afterglows of gamma-ray bursts: concurrence of jet's dynamics and wave-particle interactions}

\author{Ze-Lin Zhang}
\affiliation{School of Astronomy and Space Science, Nanjing University, Xianlin Road 163, Nanjing 210023, China}
\affiliation{Key laboratory of Modern Astronomy and Astrophysics (Nanjing University), Ministry of Education, China}

\author{Ruo-Yu Liu}
\email{ryliu@nju.edu.cn}
\affiliation{School of Astronomy and Space Science, Nanjing University, Xianlin Road 163, Nanjing 210023, China}
\affiliation{Key laboratory of Modern Astronomy and Astrophysics (Nanjing University), Ministry of Education, China}

\author{Xiang-Yu Wang}
\email{xywang@nju.edu.cn}
\affiliation{School of Astronomy and Space Science, Nanjing University, Xianlin Road 163, Nanjing 210023, China}
\affiliation{Key laboratory of Modern Astronomy and Astrophysics (Nanjing University), Ministry of Education, China}

\begin{abstract}
The origin of ultra-high energy cosmic rays $($UHECRs$)$ remains a mystery. It has been suggested that UHECRs can be produced by the stochastic acceleration in relativistic jets of gamma-ray bursts $($GRBs$)$ at the early afterglow phase. Here, we develop a time-dependent model for proton energization by cascading compressible waves in GRB jets considering the concurrent effect of the jet's dynamics and the mutual interactions between turbulent waves and particles. 
Considering fast mode of magnetosonic wave as the dominant particle scatterer and assuming interstellar medium $($ISM$)$ for the circumburst environment, our numerical results suggest that protons can be accelerated up to $\textrm{10}^{\textrm{19}}\,$eV during the early afterglow. An estimation shows ultra-high energy nuclei can easily survive photodisintegration in the external shocks in most cases, thus allowing the acceleration of $\textrm{10}^{\textrm{20}}\,$eV cosmic-ray nuclei in the proposed frame. The spectral slope can be as hard as $\ddd\NNN/\ddd\EEE \propto \EEE^\textrm{\,0}$, which is consistent with the requirement for the interpretation of intermediate-mass composition of UHECR as measured by the Pierre Auger Observatory. 
\end{abstract}


\maketitle

\section{Introduction}
\label{sec: Introduction}

Ultra-high energy cosmic rays $($UHECRs$)$ at the ankle energy $\textrm{10}^{\textrm{18.5}}\,$eV and above are the most energetic particles in nature. The presence of these particles has been known for over half a century \cite{Linsley1963}. However, the sites and mechanisms of their production are still open questions \cite{Anchordoqui2019}. The study of the energy spectrum and the mass composition of UHECR helps to reveal their origin. Recently, the results of cosmic-ray anisotropy observed by Pierre Auger Observatory and the Telescope Array support the hypothesis of an extragalactic origin for the UHECR \cite{PAO2017,TA2018}. Extragalactic sources, such as active galactic nuclei \cite{Biermann1987, Berezinsky2006}, Gamma-ray bursts $($GRBs$)$ \cite{Vietri1995,Waxman1995,Murase2006}, energetic supernovae $($such as hypernovae$)$ \cite{Wang2007,Liu2012}, tidal disruption events \cite{Farrar2009,Zhang2017,Biehl2018}, galaxy clusters \cite{Norman1995, Berezinsky1997,Vannoni2011}, as well as milli-second magnetars \cite{Arons2003, Kotera2011}, have been considered as plausible candidates of UHECR sources.

As the most powerful and intense explosive events in the universe, GRBs have been studied extensively as the cosmic accelerator of UHECRs \cite{Waxman1995,Vietri1995,Schlickeiser2000,Murase2006,Liu2011,Asano2016,Zhang2018}. However, the acceleration mechanisms of these particles in GRBs remain an enigma. The standard scenario adopted to produce non-thermal particles is the particle acceleration at shocks, e.g., diffusive shock acceleration \cite{Bell1978a,Blandford1978}. However, particle acceleration by relativistic shocks with bulk Lorentz factor $\Gamma \gg \textrm{1}$ is limited by a series of factors. For example, the relative energy gain drops quickly $($ from $\Gamma^{\textrm{2}}$ to $\simeq$ $\textrm{2}$ $)$ after the first shock crossing circle because of particles do not have sufficient time to become isotropic upstream before being caught up by the shock \cite[e.g.,][see also a recent review by \citealt{Marcowith2020}]{Gallant1999,Lemoine2006}. Another possible disadvantage of the shock acceleration is the energy budget. The required energy production rate of CRs to explain the measured flux beyond the ankle is $\textrm{10}^{\textrm{44}}\, \textrm{erg~Mpc}^{-\textrm{3}}\textrm{yr}^{-\textrm{1}}$ \cite{Katz2009,Waxman2010,Baerwald2015} while the gamma-ray energy production rate of GRBs is $\textrm{10}^{\textrm{43}}\,\textrm{erg~Mpc}^{-\textrm{3}}\textrm{yr}^{-\textrm{1}}$ for a typical gamma-ray luminosity of $\textrm{10}^{\textrm{52}}\,\textrm{erg~s}^{-\textrm{1}}$ and a local GRB rate of $\textrm{1}\,\textrm{Gpc}^{-\textrm{3}}\textrm{yr}^{-\textrm{1}}$. Given the predicted spectral slope of the accelerated particles being $\ppp\gtrsim \textrm{2}$ for the relativistic shock acceleration \cite{Bednarz1998,Achterberg2001,Lemoine2003,Keshet2005}, the fraction of the energy of CRs accelerated beyond the ankle $($$\textrm{10}^{\textrm{18.5}}\,\textrm{eV}$$)$ is only at the level of $\textrm{10}$\% of the total CR energy. As a result, it would require a baryon loading factor $($defined as the ratio of total energy populated in CRs to that in gamma rays$)$ of $\sim\,\textrm{100}$ to account for the required UHECR energy production rate. It is in tension with the constraint from the non-detection of GRB neutrinos by the IceCube neutrino telescopes in some dissipation mechanisms of GRBs \cite{IceCube2017}. Furthermore, it has been pointed out that a very hard CR injection spectrum with $\ppp\lesssim\textrm{1}$ is favored in order to fit the spectrum and composition of UHECRs measured by the Pierre Auger Observatory, where the best-fit index is even $\ppp<\textrm{0}$ \cite{Batista2019a,Batista2019b}.

Recently, a stochastic acceleration (SA) model of UHECRs via turbulence in GRB jets has been proposed to avoid the problems mentioned above \cite{Asano2016}. The SA can yield a hard UHECR spectrum with shallow index $\ppp\lesssim\textrm{2}$, which has been discussed as a possible charged particle acceleration mechanism in astrophysical plasmas \cite{Schlickeiser1984,Becker2006,Stawarz2008}. Magnetohydrodynamic (MHD) turbulence is indispensable in various astrophysical processes. As the magnetic scattering centers in SA scenario, MHD waves mainly consist of three types: incompressible Alfv\'{e}n modes and compressible fast and slow modes \cite{Cho2002}. Particle scattering and diffusion largely rely on the properties of plasma turbulence. Fast mode waves show an isotropic cascade and it could be the most effective scatterers of cosmic-rays \cite{Yan2002,Makwana2020}. The spectrum of the isotropic cascade was claimed to be $\kkk^{-\textrm{3}/\textrm{2}}$ \cite{Cho2002}.

The excitation of turbulence in plasmas generally stems from the anisotropy of particle distributions and MHD instabilities~\cite{Tsytovich1972}. In our work, we consider the turbulence driven by MHD instabilities induced by the jet's propagation in the circumburst interstellar medium $($ISM$)$, such as Kelvin-Helmholtz, Rayleigh-Taylor and Richtmyer-Meshkov instabilities~\cite{Zhang2003,Duffell2013,Matsumoto2013}. The turbulence is injected at the scale comparable to the size of the shock and then cascades down to small scales due to the wave-wave interactions. We do not emphasize any specific instability while assume the turbulent magnetic field is of the same order of the magnitude as the {}{total magnetic field $($in relativistic limit, mainly contributed by fast mode waves under our assumption\,~\cite{Makwana2020,Yan2004}$)$}. Charged particles are expected to be accelerated via the gyro-resonance with MHD waves in the condition $\omega-\kkk_{\|}v_{\|}=\textit{l}\Omega_{\textrm{g}}$ {}{$(l=\textrm{0},\pm{\textrm{1}},\pm{\textrm{2}},\dots)$}, where $\omega=\pm\kkk_{\|}v_{\textrm{w}}$ is the wave frequency, $\kkk_{\|}$ the parallel wavenumber, $v_{\textrm{w}}$ the phase velocity, and $v_{\|}=\mu v$ the particle velocity parallel to the mean magnetic field $\BBB\equiv|\textbf{\BBB}|$, and $\mu$ the pitch-angle cosine, $\Omega_{\textrm{g}}$ the gyro-frequency of relativistic particles. The positive and negative signs in the dispersion relation indicate the parallel and anti-parallel propagation of waves to $\textbf{\BBB}$. In our work, we only consider the most important resonance occurring at $\textit{l}=-\textrm{1}$ and $\kkk_{\|}=\Omega_{\textrm{g}}/v_{\|}$ , which is generally true except for $\textrm{90}^\circ$ scattering \citep{Kulsrud2005,Steinacker1992,Zhou1990}. {}{It should be noticed that gyro-resonance is not the only mechanism for wave-particle interactions in the MHD turbulence. For example, transit-time damping $($TTD, $l=\textrm{0}$ mode$)$, can also contribute to particle 
scattering especially when the pitch angle is close to $\textrm{90}^\circ$~\cite{Yan2008, Teraki2019}.}

In the previous work~\cite{Asano2016}, the authors considered the SA process with the test-particle treatment and assume non-evolving parameters such as the particle injection rate and the diffusion coefficient. In fact, acceleration of particles consumes the turbulence energy, representing as a damping process. In the meantime, it relaxes the confinement of particles in the jet and may cause particle escape from the jet. In addition, the GRB jet gets decelerated as it expands into the ISM. As a result, relevant parameters for the SA process evolves with time and particles that confined in the jet suffer the adiabatic cooling. These processes have not been considered in the previous work but they may significantly affect the SA process, and, consequently, the accelerated CR spectrum.

In this work, we attempt to model the acceleration of UHECRs via the SA process in the early afterglow of GRBs with incorporating jet's dynamics and the mutual influences between particles and the turbulence.  
  The configuration of this work proceeds as follows. In Sec.~\ref{sec: Stochastic acceleration1}, the gyro-resonant interaction of wave-particle by coupled kinetic equations in the early afterglows of GRBs is introduced. In Sec.~\ref{sec: Stochastic acceleration2}, we analysis the acceleration of UHECR by wave-particle interactions in a comprehensive way which covers the behaviors of wave-particle spectra. In Sec.~\ref{sec: Photodisintegration}, we estimate the photodisintegration rate of ultra-high energy nuclei in the external shocks under the assumption of SA. Conclusions are presented in Sec.~\ref{sec: Conclusions}. We use $\mathcal{Q}_{\textit{x}}=\mathcal{Q}/{\textrm{10}}^{\textit{\,x}}$ (i.e. $\mathcal{Q}=10^{\,\textrm{52}}\mathcal{Q}_{\textrm{52}}$, except for $\mathcal{Q}_{\textrm{300}}=\mathcal{Q}/{\textrm{300}}$) in CGS units throughout this work.

\section{Stochastic acceleration in the early afterglows of GRBs}
\label{sec: Stochastic acceleration1}
For an isotropic-equivalent, adiabatic GRB ejecta expanding in ISM~\cite{Huang1999}, the following equations have been proposed to depict its dynamic evolution \cite{Huang1999}:
  \begin{eqnarray}\label{dgamma_dmass}
  \frac{\textrm{d}\Gamma}{\textrm{d}\mmm} \simeq  -\frac{\Gamma^{\textrm{2}}-{\textrm{1}}}{\textit{M}_{\textrm{ej}}+\textrm{2}\Gamma\mmm},&\\
  \textrm{d}\mmm = \textrm{4}\pi\RRR^{\textrm{2}}\nnn_{\textrm{ISM}}\mmm_{\textrm{p}}\textrm{d}\RRR,&\\ 
  {}{\textrm{d}\RRR = \beta_{\textrm{sh}}\ccc\Gamma(\Gamma+\sqrt{\Gamma^{\textrm{2}}-\textrm{1}})\textrm{d}\ttt_{\textrm{obs}}},& 
  \end{eqnarray}
where $\Gamma$ is the bulk Lorentz factor of the external shock and the initial bulk Lorentz factor is fixed at $\Gamma_{\textrm{0}}=\textrm{300}$ in this work. $\mmm$ and $\textit{M}_{\textrm{ej}}$ are the rest mass of the swept-up ISM and the mass ejected from the GRB central engine respectively. $\RRR$ is the radius of the external shock, $\nnn_{\textrm{ISM}}$ the number density of the interstellar medium, $\mmm_{\textrm{p}}$ the mass of a proton, $\beta_{\textrm{sh}}=v_{\textrm{sh}}/\ccc$ where $v_{\textrm{sh}}$ is the bulk velocity of the material and $\ccc$ is the speed of light,~{}{ $\ttt_{\textrm{obs}}$ is the time measured in the observer's frame}.

At the onset of the afterglows $($external shocks$)$ of GRBs, the relativistic outflowing plasma can excite large scale turbulences by MHD instabilities. Particles in plasmas scatter off the randomly moving induced-turbulence, which causes the second-order Fermi acceleration~\cite{Fermi1949}. After a period of ``scattering'', the transition from anisotropic particle velocity distribution to the isotropic one $($actually, the ``scattering'' mentioned above is due to some collisionless processes between particles and fast mode waves, such as gyro-resonant wave-particle interactions~\cite{Melrose1968}$)$, hence, the reduced momentum diffusion equation can be written as~\cite{Melrose1980,Stawarz2008}:
  \begin{eqnarray}\label{DiffusiveEquation}
  \frac{\partial \fff\,(\ppp,\,\ttt)}{\partial
  \ttt}=\frac{1}{\ppp^{\textrm{2}}}\frac{\partial}{\partial\ppp}\left[\ppp^{\textrm{2}}\mathcal{D}_{\textrm{p}\textrm{p}}(
  \ppp,\ttt) \frac{\partial \fff\,(\ppp ,\ttt)}{\partial \ppp}\right],
  \end{eqnarray}
  where $f(\ppp,\ttt)$ is the phase space distribution function of momentum $\ppp$ and time $\ttt$, and $\mathcal{D}_{\textrm{p}\textrm{p}}(\ppp,\ttt)$ is the momentum diffusion coefficient which represents the rate of interaction with the turbulent fields. We adopt the energy of particle $\textit{E}$ instead of its momentum $\ppp$ by invoking $f(\ppp,\ttt)=\NNN(\textit{E},\ttt)\ddd\textit{E}/(4\pi{\ppp}^{\textrm{2}}\ddd\ppp)$. Further more, in consideration of particle injection, escape and adiabatic energy loss processes, the evolution of the proton energy distribution $\NNN(\textit{E},\ttt)$ in the outflowing plasma $($jet$)$ comoving frame can be described by the Fokker-Planck $($FP$)$ equation~\cite{Petrosian2004,Stawarz2008}:
  \begin{eqnarray}\label{KoP}
   \frac{\partial \NNN}{\partial \ttt}&=&{\frac{\partial}{\partial\textit{E}}\left[\mathcal{D}_{\textrm{E}\textrm{E}}(\textit{E},\ttt) \frac{\partial\NNN}{\partial
   \textit{E}}\right]}-{\frac{\partial}{\partial\textit{E}}
   \left[\left(\frac{{\textrm{2}}\mathcal{D}_{\textrm{E}\textrm{E}}(\textit{E},\ttt)}{\textit{E}}+\langle\,\dot{\textit{E}}\,\rangle\right)\NNN\right]}\cr\cr
   &&-\frac{\NNN}{\ttt_{\textrm{esc}}}+\mathcal{Q}_{\text{inj}}(\textit{E},\ttt),
  \end{eqnarray}
  where $\langle\,\dot{\textit{E}}\,\rangle=-\textit{E}/\ttt_{\textrm{ad}}$ represents the adiabatic energy loss of relativistic expansion, $\ttt_{\textrm{ad}}$ $=$ $\RRR/(\Gamma \ccc)$ the adiabatic energy loss timescale. The last term $\mathcal{Q}_{\textrm{inj}}(\textit{E},\ttt)=\mathcal{Q}_{\textrm{0}}(\ttt)\delta(\textit{E}-\textit{E}_{\textrm{inj}})$ represents the continuous particle injection from the initial moment, $\mathcal{Q}_{\textrm{0}}(\ttt)$ $=$ $\textrm{4}\pi\RRR^{\textrm{2}}\Gamma\nnn_{\textrm{ISM}}\ccc$ the number density at the proton injection energy $\textit{E}_{\textrm{inj}}$, and we are assuming continuous injection of particles at $\textit{E}_{\textrm{inj}}$ $=$ $\textrm{300}\,\Gamma_{\textrm{300}} \mmm_{\textrm{p}}\ccc^{\textrm{2}}$ during the early afterglows evolution, $\textit{E}=\Gamma\mmm_{\textrm{p}}\ccc^{\textrm{2}}$ the proton energy. The term $-N/{\ttt_{\textrm{esc}}}$ represents the spatial diffusive escape of the particle from the accelerated region the size of which is $\RRR/\Gamma$ in the jet's comoving frame. {}{The spatial diffusion coefficient $\mathcal{D}_{\textrm{R}\textrm{R}}$ is related to the energy diffusion coefficient $\mathcal{D}_{\textrm{E}\textrm{E}}$ by $\mathcal{D}_{\textrm{R}\textrm{R}}\mathcal{D}_{\textrm{E}\textrm{E}}=\beta_{\textrm{w}}^{\textrm{2}}\EEE^{\,\textrm{2}}$ omitting the coefficient of order unity. Therefore, the escape timescale ${\ttt_{\textrm{esc}}}={\RRR}^{\textrm{2}}/\mathcal{D}_{\textrm{R}\textrm{R}}={\RRR}^{\textrm{2}}/(\Gamma^{\textrm{2}}v_{\textrm{w}}^{\textrm{2}}\ttt_{\textrm{acc}})$~\cite{Tramacere2011}, where $\ttt_{\textrm{acc}}=\EEE^{\,\textrm{2}}/\mathcal{D}_{\textrm{E}\textrm{E}}$} the acceleration time for protons whose Larmor radii resonate with some character length scales of the turbulent magnetic fields, $v_{\textrm{\textrm{w}}}$ the phase speed of fast mode magnetosonic waves.
  The cooling effects owing to photopion production and proton synchrotron radiation can be neglected~\cite{Asano2016}. Without considering the adiabatic energy loss, we can combine the first two terms on the right-hand side of Eq.~$($\ref{KoP}$)$ into a single term $(\partial\FFF_{\textrm{p}}/\partial\textit{E})$, which represents the SA process. $\FFF_{\textrm{p}}$ can be written as
  \begin{eqnarray}\label{gain rate}
  \FFF_{\textrm{p}}(\textit{E})=\textit{E}^{\,\textrm{2}}\mathcal{D}_{\textrm{E}\textrm{E}}(\textit{E})\frac{\partial}{\partial\textit{E}}\left[\frac{\NNN (\textit{E})}{\textit{E}^{\,\textrm{2}}}\right].
  \end{eqnarray} 

Since we deal with ultra-relativstic particles, the particle velocity $v\gg v_{\textrm{\textrm{w}}}$ is considered in the numerical calculation. Hence, we use an approximated form of the diffusion coefficient in energy space given by~{}{\cite{Lynn2014,Kakuwa2016}}:
  \begin{eqnarray}
  \mathcal{D}_{\textrm{E}\textrm{E}}(\textit{E}) &\sim & \frac{\textit{E}^{\textrm{2}}\beta_{\textrm{w}}^{\textrm{2}}{\kkk_{\textrm{res}}}\ccc}{\rrr_{\textrm{g}}\uuu_{\textrm{B}}}
  \int_{{\kkk}_{\textrm{res}}}^{\kkk_{\textrm{max}}}
  {\kkk}^{-\textrm{1}}\WWW_{\textrm{B}}(\kkk){\ddd\kkk}, \label{Dee1}
  \end{eqnarray}
  where the dimensionless speed is given by~\cite{ZhangBing2018}:
  \begin{eqnarray}\label{DS}
     \beta_{\textrm{\textrm{w}}} = \frac{v_{\textrm{w}}}{\ccc} = \sqrt{\frac{\hat{\gamma}{\PPP} + \BBB^{\textrm{2}}/\textrm{4}\pi}{\rho\ccc^{\textrm{2}}+\hat{\gamma}{\PPP}/(\hat{\gamma}-1)+\BBB^{\textrm{2}}/\textrm{4}\pi}},
  \end{eqnarray} 
  and $\hat{\gamma} = \textrm{4}/\textrm{3}$ represents the adiabatic index in the relativistic regime, $\PPP=(\textrm{4}\Gamma^{\textrm{2}}\nnn_{\textrm{ISM}}\mmm_{\textrm{p}}\ccc^{\textrm{2}})/\textrm{3}$ the relativistic gas pressure, $\rho=\textrm{4}\Gamma\nnn_{\textrm{ISM}}\mmm_{\textrm{p}}$ the downstream rest mass energy density, and $\nnn_{\textrm{ISM}}$ the upstream rest number density of protons. {}{Without considering the damping effect on turbulent MHD waves, the coefficient $\mathcal{D}_{\textrm{EE}}\propto {\beta_{\textrm{w}}^{\textrm{2}}\EEE^{\textrm{2}}\ccc}/{( \rrr_{\textrm{g}}^{\textrm{2}-\qqq}\lambda_{\textrm{max}}^{\qqq-\textrm{1}})}$ is tested under different cases, such as $\qqq=\textrm{2}$ $($hard sphere approximation$)$, $\qqq=\textrm{3}/\textrm{2}$ $($Kraichnan type$)$, $\qqq=\textrm{5}/\textrm{3}$ $($Kolmogorov type$)$, and $\qqq=\textrm{1}$ $($Bohm limit$)$, and the spectral index of proton energy spectra $\EEE^{\textrm{2}}\NNN_{\textrm{CR}}(\EEE)$ are separately show $\textrm{1}$, $\textrm{3}/\textrm{2}$, $\textrm{4}/\textrm{3}$, and $\textrm{2}$, which are consistent with previous work\,\cite{Asano2016,Stawarz2008,Mertsch2011}.} The comoving magnetic field energy density will be calculated by Eq.~$($\ref{KoW}$)$ which satisfying
  \begin{eqnarray}\label{magnetic}
     \uuu_{\textrm{B}}=\frac{\BBB^{\textrm{2}}}{\textrm{8}\pi}=\int_{\kkk_{\rm \textrm{min}}}^{\kkk_{\textrm{max}}} \WWW_{\textrm{B}}(\kkk){\ddd\kkk}
  \end{eqnarray} 
  with the assumption of the magnitude of the initial magnetic field  $\BBB_{\textrm{0}}\simeq(\textrm{32}\pi\varepsilon_{\textrm{B}}\Gamma_{\textrm{0}}^{\textrm{2}}\nnn_{\textrm{ISM}}\mmm_{\textrm{p}}\ccc^{\textrm{2}})^{\textrm{1/2}}$, and $\varepsilon_{\textrm{B}}$ the magnetic field equipartition factor which indicates fraction of the magnetic field energy to the {}{internal energy $\mathcal{E}_{\textrm{tot}}$\,$($almost equal to the initial total energy of GRBs$)$. For highly turbulent plasma, we assume the energy of turbulent magnetic field is comparable to the total magnetic energy. The fast mode magnetosonic part $\WWW_{\textrm{B}}(\kkk)=\alpha\WWW(\kkk)$ being the magnetic component of the total turbulent field energy density per unit wavenumber $\WWW(\kkk)$ $($magnetic field plus plasma motion$)$. Due to the portion of fast mode waves in relativistic MHD turbulence is still not clear, here we set the dimensionless parameter $\alpha=\textrm{0.25}$~\cite{Makwana2020}. Given the turbulent energy $\mathcal{E}_{\textrm{tur}} = \varepsilon_{\textrm{T}}\mathcal{E}_{\textrm{tot}}$, where $\varepsilon_{\textrm{T}}$ is the turbulence equipartition factor.} $\kkk_{\textrm{res}}\equiv\textrm{1}/{\rrr_{\textrm{g}}}(\textit{E})$ is the corresponding wavenumber of the wave resonating with protons of energy $\EEE$, where $\rrr_{\textrm{g}}\simeq\textit{E}/(\eee\BBB)$ is the gyro-radius (Larmor radius) of the protons. $\kkk_{\textrm{min}}$ and $\kkk_{\textrm{max}}$ represent, respectively, the minimum and the maximum wavenumber of the turbulence which correspond to the injection eddy scale $\lambda_{\textrm{inj}}=\textrm{2}\pi/\kkk_{\textrm{min}}$ and the smallest eddy scale $\lambda_{\textrm{min}}=\textrm{2}\pi/\kkk_{\textrm{max}}$.  Note that the injection eddy scale $\lambda_{\textrm{inj}}$ should not be larger than the width of the shocked jet in the comoving frame $\RRR/\Gamma$ at jet's radius $\RRR$ from the central engine. Hence, we use a dimensionless parameter $\xi$ to parametrize the injection eddy scale $\lambda_{\textrm{inj}}=\xi\RRR/\Gamma\lesssim\RRR/\Gamma$. The value of {}{$\lambda_{\textrm{max}}$} is rather trivial for our calculation as long as it is smaller than the gyro-radius of protons at injection, i.e., $\textrm{2}\pi \rrr_{\textrm{g}}(\EEE_{\textrm{0}})$. We here simply set it to be $\textrm{10}^{\textrm{16}}\,$cm.

  The gyro-resonant wave-particle interactions lead to energy exchange between the turbulent waves and particles. As we mentioned above, the MHD waves in relativistic jets are taken to be isotropic, and their spectral density $\WWW(\kkk,\ttt)$ in wavenumber space is determined by the FP equation~\cite{Eichler1979,Miller1996}:
  \begin{eqnarray}\label{KoW}
  \frac{\partial \WWW}{\partial\ttt}&=&\frac{\partial}{\partial\kkk}\left[\mathcal{D}_{\textrm{k}\textrm{k}}(\kkk,\ttt) \frac{\partial \WWW}{\partial\kkk}\right] -\frac{\partial}{\partial\kkk}
  \left[\frac{{\textrm{2}}\mathcal{D}_{\textrm{k}\textrm{k}}(\kkk,\ttt)}{\kkk}\WWW\right]\cr\cr
   &&+\frac{\kkk}{\textrm{3}}(\nabla\cdot\textbf{v})\frac{\partial\WWW}{\partial\kkk} + \Gamma_{\textrm{w}}(\kkk,\ttt)\WWW + \mathcal{Q}_{\text{w, inj}}(\kkk,\ttt),
  \end{eqnarray} 
  where the third term on the RHS of the equation represents the energy loss of adiabatic expansion $($i.e.,\,$\nabla\cdot\textbf{v}>\textrm{0}$$)$ of magnetic fields at different scales and \textbf{v} the expansion velocity of the waves, $\Gamma_{\textrm{w}}(\kkk,\ttt)$ represents the damping effect, and $\mathcal{Q}_{\textrm{w, inj}}(\kkk,\ttt)=\mathcal{Q}_{\textrm{w0}}(t)\delta(\kkk-\kkk_{\textrm{inj}})$ represents the continuous energy injection into the turbulence at a mono-scale $\lambda_{\textrm{inj}}=\textrm{1}/\kkk_{\textrm{inj}}$, where $\mathcal{Q}_{\textrm{w}\textrm{0}}$ $=$ {}{ $\textrm{4}\Gamma^{\textrm{2}}\varepsilon_{\textrm{T}}\nnn_{\textrm{ISM}}\mmm_{\textrm{p}}\ccc^{\textrm{2}}/(\RRR/\Gamma\ccc)$} the injection rate per unit volume at the wavenumber $\kkk_{\textrm{inj}}$.  As the jet's expansion, $\kkk_{\textrm{inj}}$ will gradually get smaller. Note that $\kkk_{\textrm{inj}}$ is not to be confused with another characteristic wavenumber $\kkk_{\textrm{res, inj}}$, which corresponds to the wave resonate with protons at injection energy $\EEE_{\textrm{inj}}$. 
  
  \begin{figure*}[t]
  \includegraphics[width=7in]{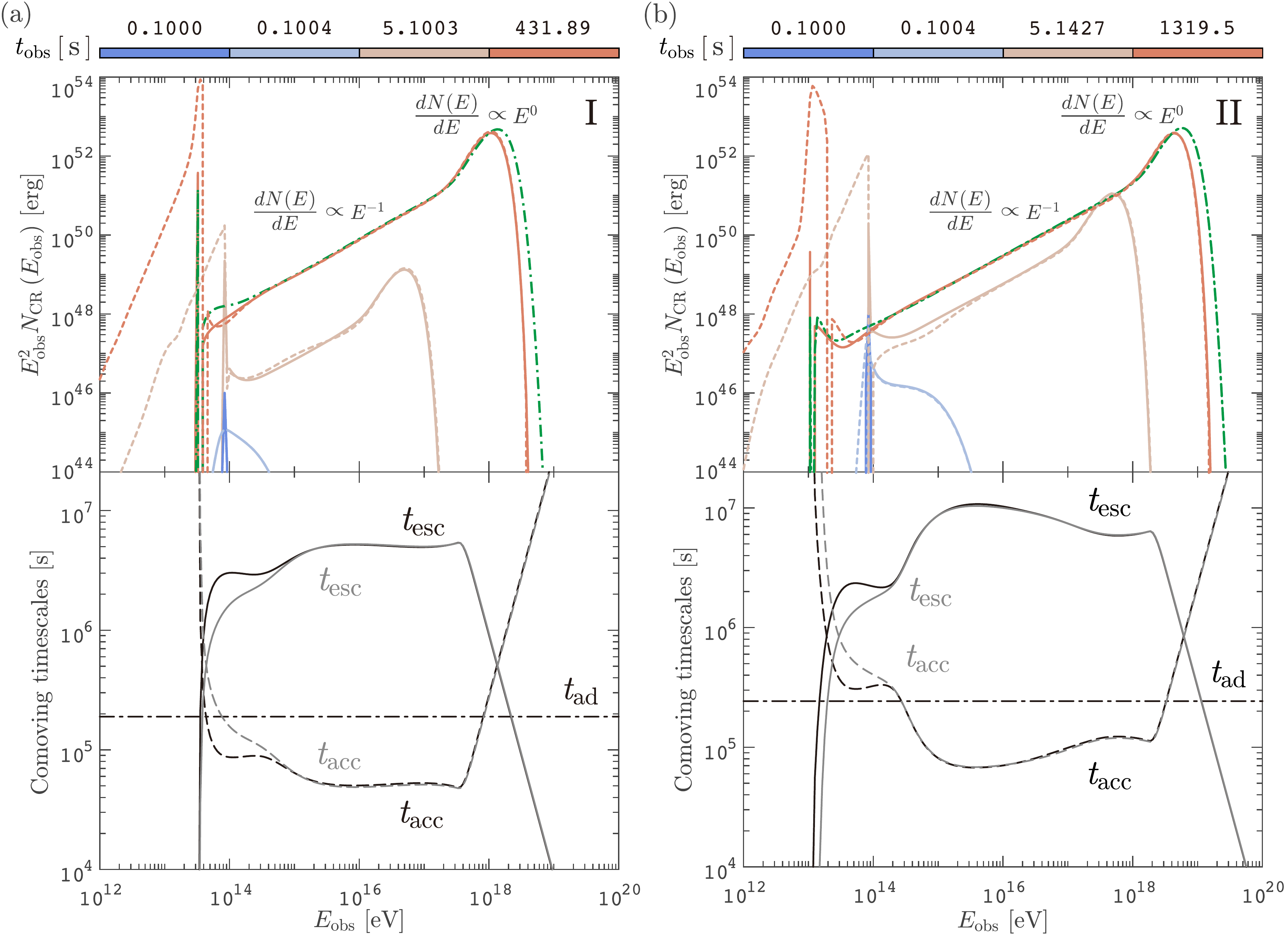}
  \caption{UHECR protons spectra resulting from joint stochastic acceleration, particle diffusive escape, and adiabatic energy loss and their corresponding comoving timescales as a function of observed proton energy. {Upper panels}: The dark blue lines represent proton injection at $\textrm{0.1000}$ s in observer's frame. The colored solid lines from dark blue to dark red represent the evolution of proton spectrum under the case of $\xi=\textrm{0.1}$, respectively. The corresponding colored short-dashed lines depict the evolution without considering the particle spatial diffusive escape effect. The green dash-dotted lines in the upper panels delineate the cases ignoring the adiabatic energy loss at the final moment of the evolution. {Lower panels}: Comoving timescales against observed proton energy under the case of $\xi=\textrm{0.1}$. The evolution of the energy spectra of protons for $\textrm{100000}$\,s in the comoving frame of relativistic outflowing plasma with $\nnn_{\textrm{ISM}}=\textrm{0.01}~\textrm{cm}^{-\textrm{3}}$ $($case \textbf{I}$)$ and $\nnn_{\textrm{ISM}}=\textrm{1}~\textrm{cm}^{-\textrm{3}}$ $($case \textbf{II}$)$. {}{The acceleration, adiabatic expansion cooling and diffusive escape timescales are separately denoted by dashed, dash-dotted and solid lines}. The gray lines show the evolution without including adiabatic cooling process. Proton spectra evolve during: {}{$($$\textrm{a}$$)$ $\ttt_{\textrm{obs}}\in[{\textrm{0.1000}\textrm{s}},~{\textrm{431.89}\,\textrm{s}}]$ and $($$\textrm{b}$$)$ $\ttt_{\textrm{obs}}\in[{\textrm{0.1000}\textrm{s}},~{\textrm{1319.5}\,\textrm{s}}]$ in the observer's frame, respectively.} We only show the final moment of different timescales in the lower panels. {}{More moments of the evolution of proton spectra are shown in Appendix.\,\ref{sec:Skeleton}.}}
  \label{fig:Fig1} 
\end{figure*} 
  
  The first two terms on the right-hand side of Eq.~$($\ref{KoW}$)$ indicate the energy cascade process in the wavenumber space, which can be reformulated into the same form of Eq.~$($\ref{gain rate}$)$ as $\kkk^{\textrm{2}}\mathcal{D}_{\textrm{k}\textrm{k}}(\kkk)\partial/\partial\kkk\left[\WWW (\kkk)/\kkk^{\textrm{2}}\right]$.


 Since we consider the compressible fast mode waves, the Iroshnikov-Kraichnan-type (IK-type) turbulence is adopted, and the diffusion coefficient in wavenumber space 
  $\mathcal{D}_{\textrm{k}\textrm{k}}(\kkk)$ can be given by~\cite{Miller1996}:
  \begin{eqnarray}\label{DKK}
   \mathcal{D}_{\textrm{k}\textrm{k}}(\kkk) = \mathcal{C}^{\textrm{2}}\kkk^{\textrm{4}}v_{\textrm{w}}\bigg[\dfrac{\WWW\left(\kkk\right)}{\textrm{2}\textit{u}_{\textrm{B}}}\bigg],
  \end{eqnarray}
  where $\mathcal{C}$ is the {}{Kolmogorov} constant of order unity. 
  Note that turbulence could be already driven in the jet before the onset of the afterglow phase (i.e., during the prompt emission phase), so we assume an initial condition for $\WWW(\kkk,\ttt)$ as  
  \begin{eqnarray}\label{turbulent injection}
  \WWW(\kkk,\ttt=0) \equiv {}{\kappa_{\textrm{0}}\uuu_{\textrm{T}}\left(\frac{\kkk}{\kkk_{\textrm{inj}}}\right)^{-\qqq}
  \textrm{exp}\left(-\frac{\kkk}{\kkk_{\textrm{max}}}\right)},
  \end{eqnarray}
  where the parameter $\kappa_{\textrm{0}} \approx -{\textrm{2}}\kkk_{\textrm{inj}}^{\,\qqq}\left(\kkk_{\textrm{max}}^{-\qqq+{\textrm{1}}}-\kkk_{\textrm{inj}}^{-\qqq+{\textrm{1}}}\right)$, {}{$\uuu_{\textrm{T}}=\textrm{4}\Gamma^{\textrm{2}}\varepsilon_{\textrm{T}}\nnn_{\textrm{ISM}}\mmm_{\textrm{p}}\ccc^{\textrm{2}}$ the comoving turbulent field energy density}, and the IK-type spectral index $\qqq = {\textrm{3}}/\textrm{2}$. Note that the damping effect, if not negligible, would cause the deviation of the turbulence spectrum from the IK spectrum.

  The energy gain of particles serves as a damping process for the turbulence. We here only consider the damping of the turbulence due to the gyro-resonance of protons. Therefore, the energy dissipation rate of the {}{turbulence} should be equal to the energy gain rate of the protons~\cite{Brunetti2007}, i.e., 
  \begin{eqnarray}\label{damping rate}
  \int\ddd\kkk~\Gamma_{\textrm{w}}(\kkk)\WWW(\kkk) = -\int\ddd\textit{E}~\textit{E}\frac{\partial\FFF_{\textrm{p}}(\textit{E})}{\partial\textit{E}}.
  \end{eqnarray}

  From Eq.~$($\ref{Dee1}$)$, integrating by parts twice, we obtain
  \begin{eqnarray}\label{damping}
  \Gamma_{\textrm{w}}(\kkk) = -\frac{{\textrm{4}}\pi \eee^{\textrm{2}} \beta_{\textrm{w}}^{\textrm{2}}
  \ccc}{\kkk}\bigg[{}{\nnn\big(\textit{E}_{\textrm{res}}(\kkk)\big)}+\int_{\textit{E}_{\textrm{res}}
  (\kkk)}^{\textit{E}_{\textrm{max}}}\frac{{\textrm{2}}{}{\nnn(\textit{E})}}{\textit{E}}\ddd\textit{E}\bigg],~~
  \end{eqnarray}
{}{where $\nnn(\textit{E})\equiv{\NNN(\textit{E} )}/{\textit{V}}$ represents the number density and the volume of the acceleration zone in the jet's comoving frame is estimated by $\textrm{}\textit{V}=\textrm{4}\pi\RRR^{\textrm{2}}\cdot\RRR/\Gamma$.} The turbulence at the wavenumber $\kkk$ is damped by protons with energy $\textit{E}>\textit{E}_{\textrm{res}}(\kkk)$ where $\EEE_{\textrm{res}}=\eee\BBB/\kkk$. The turbulent magnetic fields in the relativistic jet indicate $\delta\BBB\lesssim\BBB$.

\begin{figure*}[t]
  \includegraphics[width=7in]{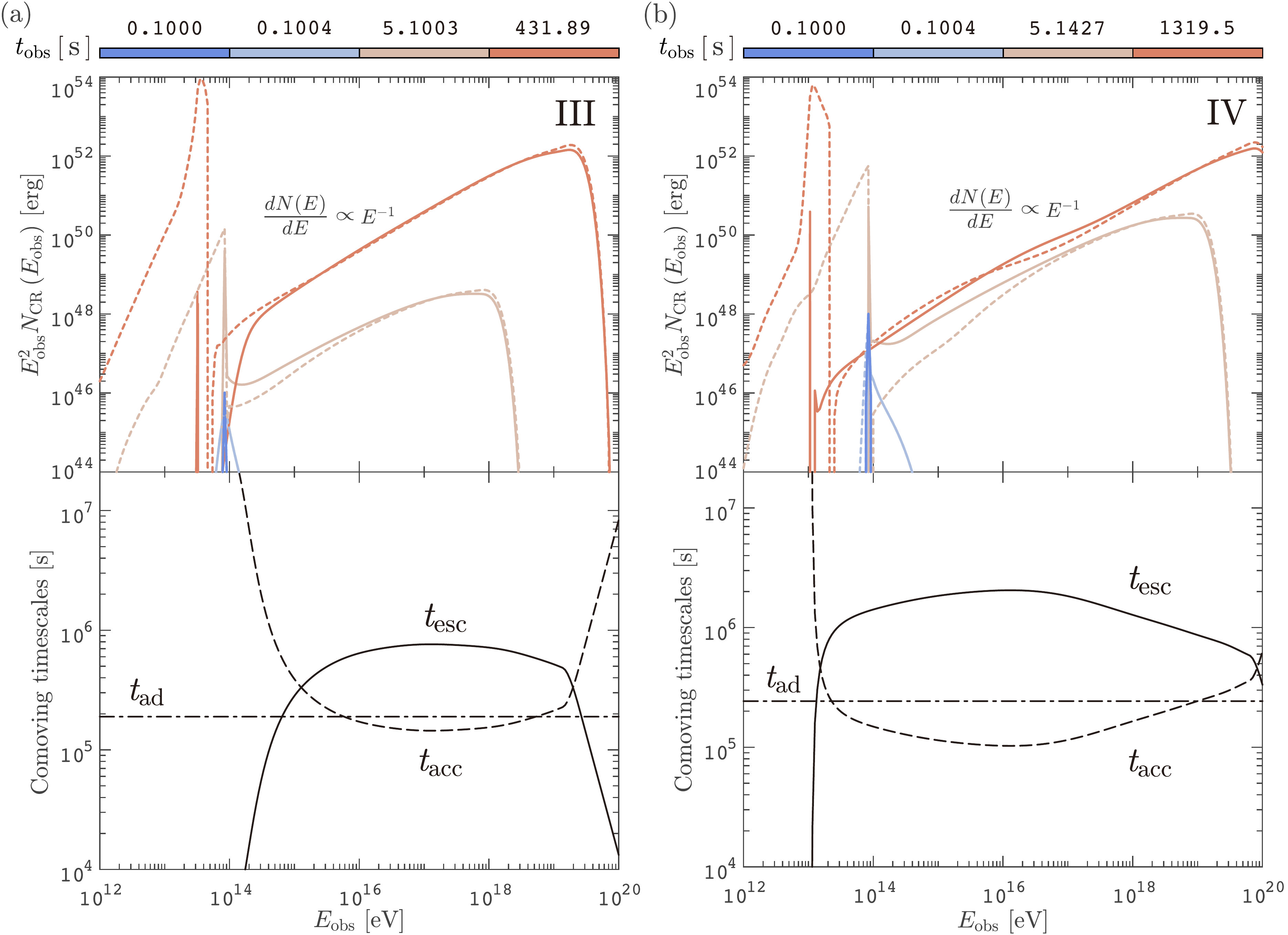}
  \caption{Consistent with the descriptions in Fig.\,\ref{fig:Fig1}. {Upper panels}: {}{The lines which from dark blue to dark red represent the evolution of proton spectrum under} the case of $\xi={\textrm{1}}$, respectively. {Lower panels}: Comoving timescales against observed proton energy under the case of $\xi={\textrm{1}}$. The evolution of the energy spectra of protons for $\textrm{100000}$\,s in the comoving frame of relativistic outflowing plasma with $\nnn_{\textrm{ISM}}=\textrm{0.01}\,\textrm{cm}^{-\textrm{3}}$ $($case \textbf{III}$)$ and $\nnn_{\textrm{ISM}}=\textrm{1}~\textrm{cm}^{-\textrm{3}}$ $($case \textbf{IV}$)$. {}{For simplicity, here we just compare the cases with and without the escape effect. More moments of the evolution of proton spectra are shown in Appendix.\,\ref{sec:Skeleton}.}}
  \label{fig:Fig2}
\end{figure*}

\begin{figure*}[t]
  \includegraphics[width=6.6in]{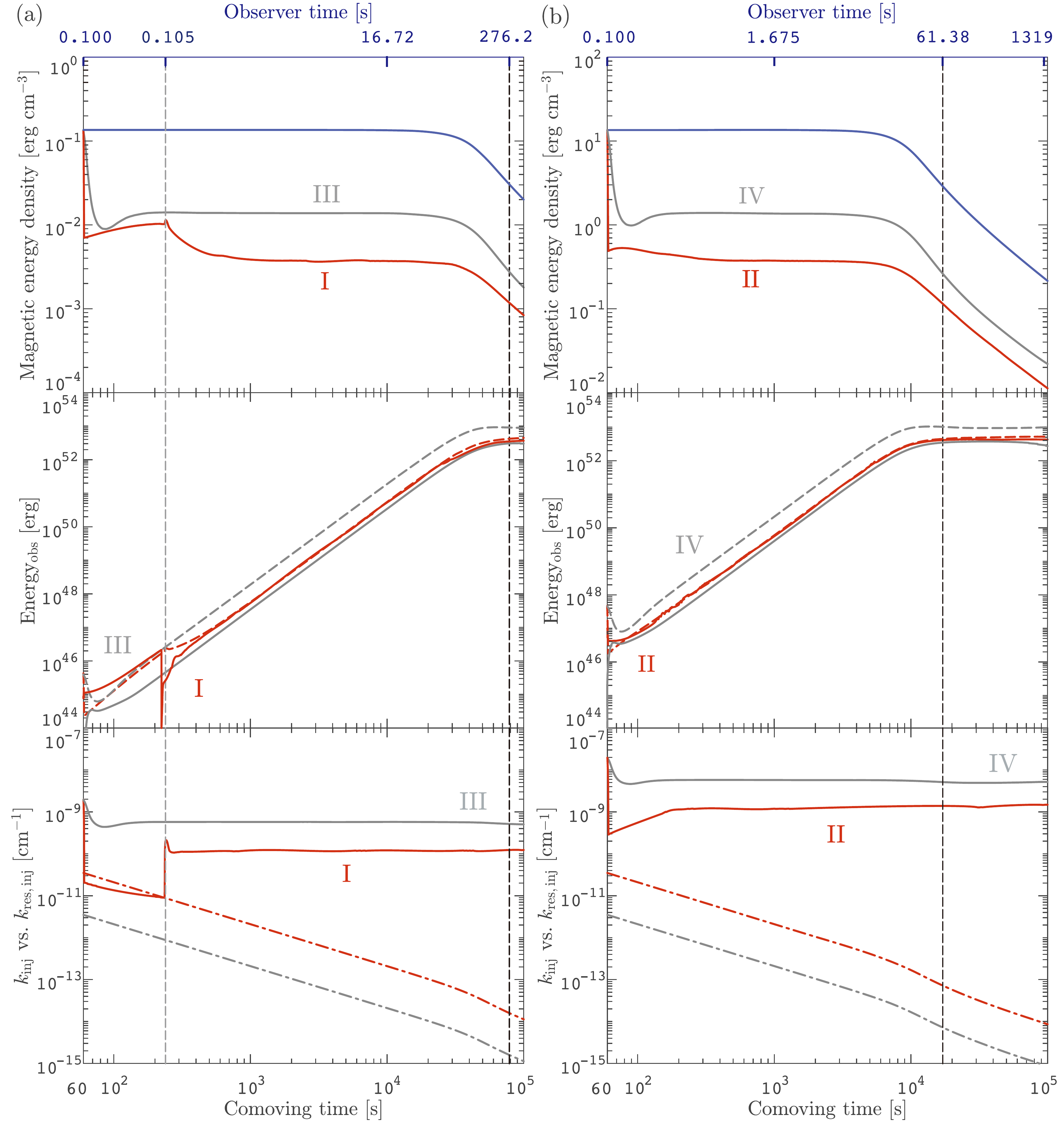}
  \caption{{}{{Top panels}: The evolution of magnetic field energy density in the phase of GRB's early afterglows which are under the four different cases. The blue solid lines represent the evolution of the magnetic fields energy density in GRB blast wave $($the downstream of the shocked fluid$)$ without considering wave-particle interactions calculated by  $\BBB\simeq\sqrt{\textrm{32}\pi\varepsilon_{\textrm{B}}\Gamma^{\textrm{2}}\nnn_{\textrm{ISM}}\mmm_{\textrm{p}}\ccc^{\textrm{2}}}$. $($$\textrm{a}$$)$ $\nnn_{\textrm{ISM}}=\textrm{0.01}\,{\textrm{cm}}^{-\textrm{3}}$ and $($$\textrm{b}$$)$ $\nnn_{\textrm{ISM}}=\textrm{1}\,{\textrm{cm}}^{-\textrm{3}}$. The colored solid lines $($gray for $\xi=\textrm{1}$ and red for $\xi=\textrm{0.1}${}$)$ show the evolution of the magnetic energy density components $($without fast mode waves which are consumed by protons$)$ of the turbulent magnetic fields with damping by protons  are calculated by the Eq.~$($\ref{magnetic}$)$. Black dashed lines represent deceleration time. {Middle panels}: The evolution of turbulence energy $($dashed lines$)$ and cosmic-ray energy $($solid lines$)$ in 
 the observer's frame. {Bottom panels}: The injection wavenumber $\kkk_{\textrm{inj}}$ $($dash-dotted lines$)$ and resonant injection wavenumber $\kkk_{\textrm{res,\,inj}}$ $($solid lines$)$. The vertical gray dashed line represents the moment when $\kkk_{\textrm{inj}}=\kkk_{\textrm{res,\,inj}}$ again.}}
  \label{fig:Fig3}
\end{figure*}

\section{Results and discussions of turbulent stochastic acceleration}
\label{sec: Stochastic acceleration2}
We adopt the {}{Runge}-Kutta method to solve the dynamical evolution of the GRB jet, and the central difference method to solve the time-dependent FP equations, see details in the Appendix of Ref.~\cite{Liu2017}. UHECR protons accelerated by turbulence through wave-particle gyro-resonant interactions are considered under four different cases, ``\textbf{I}'' for $\xi=\textrm{0.1}$ and $\nnn_{\textrm{ISM}}=\textrm{0.01}\,{\textrm{cm}}^{-\textrm{3}}$, ``\textbf{II}'' for $\xi=\textrm{0.1}$ and $\nnn_{\textrm{ISM}}=\textrm{1}\,{\textrm{cm}}^{-\textrm{3}}$, ``\textbf{III}'' for $\xi=\textrm{1}$ and $\nnn_{\textrm{ISM}}=\textrm{0.01}\,{\textrm{cm}}^{-\textrm{3}}$ and ``\textbf{IV}'' for $\xi=\textrm{1}$ and $\nnn_{\textrm{ISM}}=\textrm{1}\,{\textrm{cm}}^{-\textrm{3}}$. All these cases take the initial bulk Lorentz factor $\Gamma_{\textrm{0}}=\textrm{300}\,\Gamma_{\textrm{300}}$ and consider the onset of the afterglow at $\textrm{0.1}\,$s $($in the observer's frame$)$ after the burst so that the initial radius of the early afterglows are set to $\RRR_{\textrm{0}}\simeq\,\textrm{2}\Gamma_{\textrm{0}}^{\textrm{2}}\ccc\ttt_{\textrm{0}}=\textrm{5.4}\times\textrm{10}^{\textrm{14}}~\textrm{cm}$. The time-dependent proton spectra in these cases are shown in the upper panels of Fig.\,\ref{fig:Fig1} and Fig.\,\ref{fig:Fig2}, where some relevant timescales are shown in the lower panels. {}{Note that here we just show the spectra in four typical moments. We present the spectra at more moments of the evolution in Appendix.\,\ref{sec:Skeleton}.}

By comparing the final spectra of the four cases, we can see that the maximum accelerated energy is roughly proportional to $\xi$, implying that the particle acceleration in the early afterglow is mainly limited by the eddy size or the longest wavelength of the turbulence. This agrees with the result in Ref.~\cite{Asano2016}. Considering the adiabatic cooling slightly softens the spectrum at the cutoff regime (where $\ttt_{\textrm{acc}}\simeq \ttt_{\textrm{ad}}$ or $\EEE\simeq \EEE_{\textrm{eq}}$) as shown with the {}{green dash-dotted} lines. A smaller $\xi$, on the other hand, leads to a hardening or a pile-up spectral feature at the high-energy end. This is because the same energy of turbulence would then distribute over a narrower span in the wavenumber space given a smaller $\xi$, and hence enhances the energy density in per unit wavenumber (i.e., a larger $\mathcal{D}_{\textrm{EE}}$). As a consequence, the SA process would push protons to higher energy more efficiently, and on the other hand, a smaller eddy size result in the termination of wave-particle interactions at smaller energy. These two effects jointly lead to the formation of the pile-up bump. Diffusive escape of particles does not have significant influence on the spectrum at the high-energy end, but play an important role in shaping the spectrum around $\EEE_{\textrm{inj}}$, as shown in the upper panels of Fig.\,\ref{fig:Fig1} and Fig.\,\ref{fig:Fig2}. {}{The eddies around the resonant injection scale $\sim\textrm{1}/\kkk_{\textrm{res,\,inj}}$ are largely consumed by the injected particles. In the meanwhile, the number of scatterers $($eddies$)$ drops quickly, particles can no longer be bound by waves. Therefore, particles can efficiently escape from the present acceleration region, causing the reduction of the number of particles in the acceleration zone, as three cases $($\textbf{II}, \textbf{III} and \textbf{IV}$)$ shown in Appendix.\,\ref{sec:number}, while the specificity of case \textbf{I} will be discussed separately below.} This can be also seen by comparing the timescales shown in the lower panels in Fig.\,\ref{fig:Fig1} and Fig.\,\ref{fig:Fig2}. At the high energy end, when the acceleration timescale becomes comparable to the adiabatic cooling timescale (which is also comparable to the dynamical timescale), the diffusive escape timescale is still several times longer. From Fig.\,\ref{fig:Fig1}, we can see that the influence of the adiabatic cooling effect to the spectrum of the proton is not significant. As shown in the top panels of Fig.\,\ref{fig:Fig1}, it is worth noting that, the total kinetic energy (or thermal energy in the rest frame if assuming swept-up protons are isotropized in the downstream of the shock) of protons at injection is $\mathcal{E}_{\textrm{tot}}\sim \Gamma^{\textrm{2}}\textit{M}_{\textrm{sw}}\ccc^{\textrm{2}}\sim \textrm{10}^{\textrm{54}}$\,ergs where $\textit{M}_{\textrm{sw}}$ is the mass of swept-up material, but protons are accelerated via extracting the turbulent magnetic field energy and hence the total proton energy is restricted by the magnetic equipartition factor {}{$\varepsilon_{\textrm{B}}=\alpha\varepsilon_{\textrm{T}}$. As a result, the baryon loading factor of accelerated protons is naturally determined instead of manual selection. It should be noted that in order to ensure the validity of UHECR acceleration above the ankle in our model, the value of $\varepsilon_{\textrm{B}}$ should not be much less than $\textrm{0.1}$. For a local GRB rate of $\textrm{1}\,\textrm{Gpc}^{-\textrm{3}}\textrm{yr}^{-\textrm{1}}$, the required cosmic-ray energy budget should be about $\textrm{10}^{\textrm{53}}\,\textrm{erg}$, given the inferred CR energy production rate of $\textrm{10}^{\textrm{44}}\,\textrm{erg}\,\textrm{Mpc}^{-\textrm{3}}\textrm{yr}^{-\textrm{1}}$. For GRBs with a typical total kinetic energy  $\mathcal{E}_{\textrm{tot}}=\textrm{10}^{\textrm{54}}\,\textrm{ergs}$, it would be insufficient to explain the origin of UHECRs with SA if $\varepsilon_{\textrm{B}}\ll\textrm{0.1}$.}

Comparing Fig.\,\ref{fig:Fig1} with Fig.\,\ref{fig:Fig2}, we observe that the maximum energy is also related to the ambient ISM density. At the early afterglow phase, the jet has not been significantly decelerated so that the difference of the bulk Lorentz factor $\Gamma$. Therefore, the turbulence energy injection rate $\mathcal{Q}_{\textrm{w, inj}}$ is proportional to the ambient gas density. A higher ISM density converts more kinetic energy into the magnetic energy, and hence a larger diffusion coefficient, which facilitates the acceleration, can be expected.

\begin{figure*}[t]
  \includegraphics[width=7in]{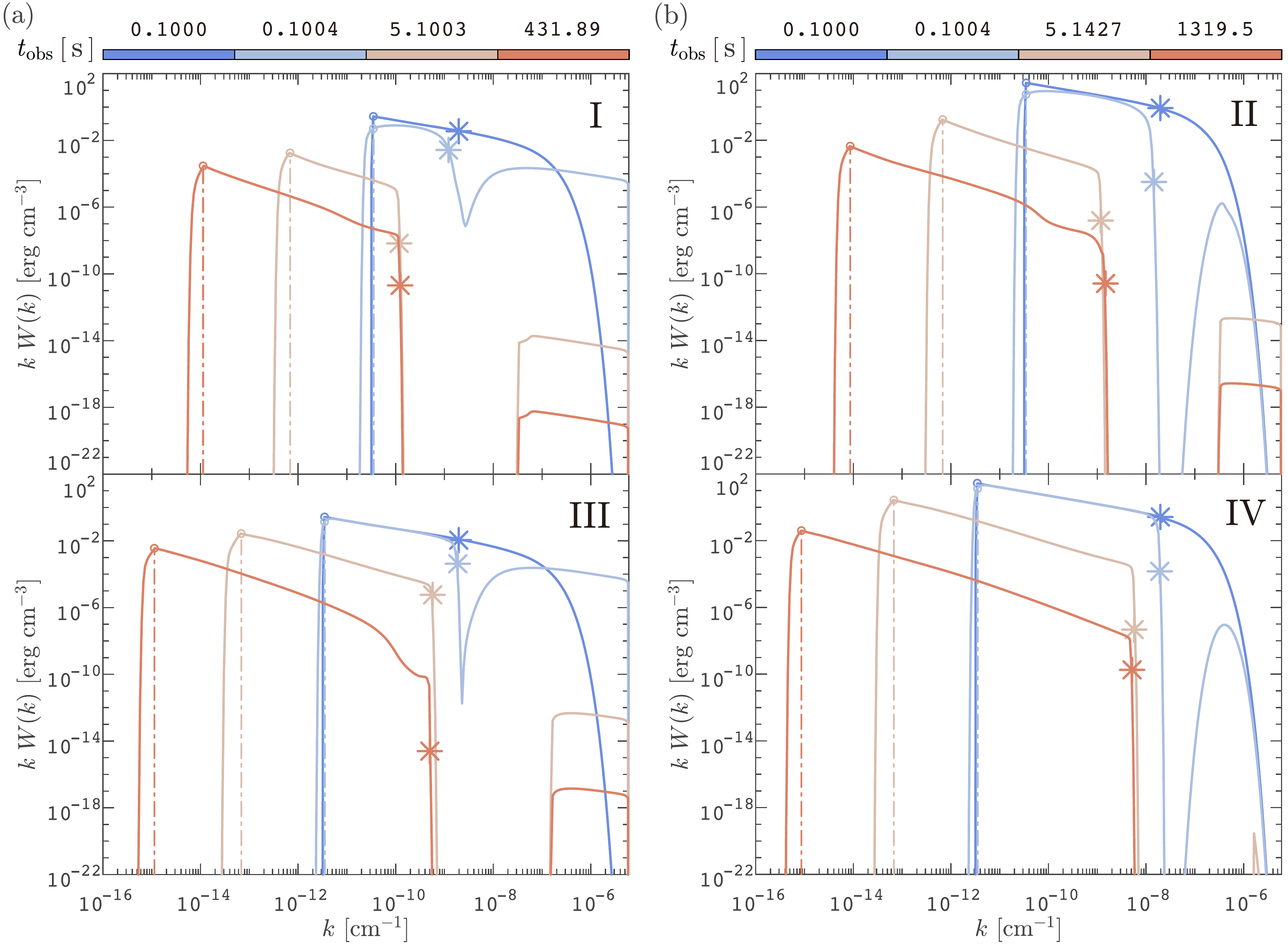}
  \caption{Turbulence spectral energy density against wavenumber. {}{The initial wave energy $($magnetic field plus plasma motion$)$ injection at $\textrm{0.1000}$ s in observer's frame are separately represented by the dark blue solid lines in upper panels $($$\xi=\textrm{0.1}$$)$ and lower panels $($$\xi=\textrm{1}$$)$.} The solid lines from dark blue to dark red represent the evolution of the turbulent waves under the case of $\xi=\textrm{0.1}$ $($case \textbf{I} and case \textbf{II}$)$ and $\xi=\textrm{1}$ $($case \textbf{III} and case \textbf{IV}$)$, respectively. The evolution of the relativistic outflowing plasma wave spectra for $\ttt_{\textrm{cmv}}=\textrm{100000}$\,s with $\nnn_{\textrm{ISM}}=\textrm{0.01}~\textrm{cm}^{-\textrm{3}}$ $($left panels$)$ and $\nnn_{\textrm{ISM}}=\textrm{1}~\textrm{cm}^{-\textrm{3}}$ $($right panels$)$. Evolution time: $($a$)$ $\ttt_{\textrm{obs}}\in[{\textrm{0.1000}\,\textrm{s}},~{\textrm{431.89}\,\textrm{s}}]$ and $($b$)$ $\ttt_{\textrm{obs}}\in[{\textrm{0.1000}\textrm{s}},~{\textrm{1319.5}\,\textrm{s}}]$ in the observer's frame, respectively. Asterisks on the lines show the turbulent magnetic field energy density which are corresponding to the wavenumber $\kkk_{\textrm{res,\,inj}}$. The circles at different moments show the injected position of turbulent waves. The dash-dotted lines represent the cut-off position of injection wavenumber at different moments. {}{More moments of the turbulence spectra are shown in Appendix.\,\ref{sec:Skeleton}.}}
  \label{fig:Fig4}
\end{figure*}

According to Ref.\,\cite{Stawarz2008}, if $\qqq=\textrm{3}/\textrm{2}$, the steady-state particle spectrum implied by Eq.\,$($\ref{KoP}$)$ is $\ddd\NNN/\ddd\EEE\propto\EEE^{\,\textrm{1}-\textit{q}}$ when $\EEE\in(\EEE_{\textrm{inj}},\,\EEE_{\textrm{eq}})$, as long as the particle escape can be neglected $($$\ttt_{\textrm{esc}}\gg\ttt_{\textrm{acc}}$, $\ttt_{\textrm{ad}}$$)$. So the power-law energy spectra $\EEE^{\textrm{\,2}}\NNN_{\textrm{CR}}(\EEE)$ is proportional to $\EEE^{\,\textrm{3}/\textrm{2}}$. This is the result obtained in the test particle limit and without considering the dynamic evolution of the system. From Fig.\,\ref{fig:Fig1} and Fig.\,\ref{fig:Fig2}, we see that the bulk of the accelerated particle spectra in all four considered cases are softer. In general, when taking into account the feedback of particle acceleration on the turbulence, the turbulence energy is consumed. Such a negative feedback from the protons impedes themselves to be further accelerated. The feedback is also reflected in the magnetic field strength, as can be seen from Fig.\,\ref{fig:Fig3} where we compare the evolution of the magnetic field under the feedback with that expected in the standard GRB afterglow dynamic model. It is interesting to note that many previous literature found a very small $\varepsilon_{\textrm{B}}$ for the external shock when modelling the multi-wavelength afterglow of some GRBs (e.g., Ref.\cite{Kumar10, Liu11, Lemoine13}), which significantly deviates from the energy-equipartition value. We speculate that the feedback of the particle acceleration on the turbulence energy could be a reason. This will be studied elsewhere. 

To show the tendency of energy transfer from turbulent magnetic field to particles, we compare the magnetic field energy density evolution under the four different cases, as shown in Fig.\,\ref{fig:Fig3}. {}{Since the escape effect is considered in our model, the UHECR spectrum should be based on the escaped particles. If protons are confined in the shocked region, the protons lose energy via adiabatic cooling. The evolution time of the final UHECR spectra escaped from the region should longer than the deceleration time $\ttt_{\textrm{dec}}$. We know that the GRB jets start decelerating at a typical radius,
\begin{eqnarray}\label{rdec}
\RRR_{\textrm{dec}}&\equiv&\left(\frac{\textrm{3}\EEE_{\textrm{tot}}}{\textrm{4}\pi\Gamma_{\textrm{0}}^{\textrm{2}}\nnn_{\textrm{ISM}}\mmm_{\textrm{p}}\ccc^{\textrm{2}}}\right)^{\textrm{1/3}}\cr\cr
&\simeq&
	\textrm{1.2}\times\textrm{10}^{\textrm{17}}\nnn_{\textrm{ISM}}^{-\textrm{1/3}}\left(\frac{\displaystyle\EEE_{\textrm{tot}}}{\displaystyle\textrm{10}^{\textrm{54}}\,\textrm{erg}}\right)^{\textrm{1/3}}\left(\frac{\displaystyle\Gamma_{\textrm{0}}}{\displaystyle\textrm{300}}\right)^{-\textrm{2/3}}\,\textrm{cm},~~~~~~
\end{eqnarray}
therefore, in the cases \textbf{I} and \textbf{III}, the deceleration radius $\RRR_{\textrm{dec}}\simeq\textrm{5.6}\times\textrm{10}^{\textrm{17}}\,{\textrm{cm}}$, and about $\textrm{1.2}\times\textrm{10}^{\textrm{17}}\,{\textrm{cm}}$ in the cases \textbf{II} and \textbf{IV}. Their corresponding deceleration timescales are $\ttt_{\textrm{dec}}\simeq\textrm{7.8}\times\textrm{10}^{\textrm{4}}\,\textrm{s}$ and $\textrm{1.7}\times\textrm{10}^{\textrm{4}}\,\textrm{s}$ in jet's comoving frame. To ensure the final UHECR spectra escaped from the region after calculations longer than $\textrm{t}_{\textrm{dec}}$, we set the evolution timescale of the wave-particle system in jet's comoving frame $\ttt_{\textrm{cmv}}=\textrm{1.0}\times\textrm{10}^{\textrm{5}}~\textrm{s}$, as shown in Fig.\,\ref{fig:Fig3}.}

{}{The evolution of turbulence energy and cosmic-ray energy are shown in the middle panels of Fig.\,\ref{fig:Fig3}. The turbulence energy is calculated by 
\begin{eqnarray}\label{VW}
\mathcal{E}_{\textrm{tur}}=\Gamma\VVV\int_{\kkk_{\textrm{min}}}^{\kkk_{\textrm{max}}}\WWW(k)\ddd\kkk,
\end{eqnarray}
and the corresponding cosmic-ray energy is given by
\begin{eqnarray}\label{ECR}
\mathcal{E}_{\textrm{CR}}=\Gamma\int_{\EEE_{\textrm{acc}}}^{\EEE_{\textrm{max}}}\EEE\frac{\ddd\NNN(\EEE)}{\ddd\EEE}\ddd\EEE
\end{eqnarray}
in the observer's frame. Here, $\EEE_{\textrm{acc}}$ is adopted by $\textrm{5}\EEE_{\textrm{inj}}$ for all cases. In cases \textbf{III} and \textbf{IV}, we can see that $\mathcal{E}_{\textrm{tur}}$ is almost $\textrm{10}\,\%$ of $\mathcal{E}_{\textrm{tot}}$ $($$\varepsilon_{\textrm{T}}=\textrm{0.1}$$)$ in equilibrium state, and the cosmic-ray energy which is extracted from the fast mode waves energy is about $\textrm{25}\,\%$ to $\mathcal{E}_{\textrm{tur}}$ $(${}$\alpha=\textrm{0.25}${}$)$. However, from cases \textbf{I} and \textbf{II}, we can see that $\mathcal{E}_{\textrm{tur}}$ is only a slightly larger than $\mathcal{E}_{\textrm{CR}}$. Due to the smaller $\xi=\textrm{0.1}$ in cases \textbf{I} and \textbf{II}, the position of resonant injection wavenumber $\kkk_{\textrm{res,\,inj}}$ is more closer to the injection wavenumber $\kkk_{\textrm{inj}}$ $($higher energetic waves$)$, as shown in cases \textbf{I} and \textbf{II} of Fig.\,\ref{fig:App1} in Appendix.\,\ref{sec:Skeleton}. We noticed that the magnitude of the turbulent magnetic fields in cases \textbf{III} and \textbf{IV} $($the conservation of turbulent magnetic field are well maintained$)$ are almost twice higher than that in cases \textbf{I} and \textbf{II} during the evolution. It means protons get twice as much energy in cases \textbf{I} and \textbf{II} than in cases \textbf{III} and \textbf{IV}. With fewer scatterers exist, these energized protons are more likely to escape from the acceleration region, and it is, as shown in Fig.\,\ref{fig:App2} in Appendix.\,\ref{sec:number}. The turbulence energy in cases \textbf{I} and \textbf{II} is half of that in cases \textbf{III} and \textbf{IV}, the reason is that damping of waves even occurs at the wave injection scale $\sim\kkk_{\textrm{inj}}$.}

{}{In addition, we noticed that the complex behavior of case \textbf{I} is related to the relative values between $\kkk_{\textrm{inj}}$ and $\kkk_{\textrm{res,\,inj}}$. The detailed explanation of it is given in Appendix.\,\ref{sec:vs}}.

In the meanwhile, our model requires that the wavenumber $\kkk$ should not be less than the injection wavenumber $\kkk_{\textrm{inj}}$, as shown in the final moment of evolution of Fig.\,\ref{fig:Fig4}.  However, the diffusive nature of FP equation allow the existence of smaller wavenumbers than $\kkk_{\textrm{inj}}$. In our calculation of turbulent magnetic field, the distribution of wavenumbers $\kkk<\kkk_{\textrm{inj}}$ is omitted. Hence, we get a relatively small value of turbulent magnetic field under the case of $\xi=\textrm{0.1}$.

Furthermore, the magnetic energy is also lost due to the adiabatic expansion of the jet. Since we assume the injection eddy size to be proportional to the jet's radius, the expansion of the jet also reduces the injection wavenumber of the turbulence $\kkk_{\textrm{inj}}$. The turbulent energy would then distribute over a larger and larger range in the wavenumber space, so that the energy density per wavenumber is reduced. Therefore, compared to the case in the test particle limit and the steady state, there will be a decline in the capacity of the stochastic acceleration with time. This is also reflected in the particle spectrum. We can see the bulk of the accelerated particle spectrum is softer than $\EEE^{\,\textrm{3}/\textrm{2}}$. 


According to above parameters evolution, the shape of the wave energy density spectra can be easily settled down from two types of wavenumber, $\kkk_{\textrm{inj}}$ and $\kkk_{\textrm{res, inj}}$, {}{as shown in Fig.\,\ref{fig:Fig3} and Fig.\,\ref{fig:Fig4}}. As the turbulent eddy scale becomes larger and larger, the wavenumber of the it becomes smaller and smaller. The larger wavenumber associated eddies $($smaller {}{scales}$)$ have already been damped by the corresponding lower energy particles, the relative higher energy particles trapped in the acceleration region which can continuously gain energy from the lower wavenumber turbulent waves. Then energy transport in $\kkk$-space will cause more remarkable deviation from the IK-type spectrum in lower wavenumber.

In the case of ISM environment around bursts, our results suggest that a combination of cyclotron wave damping and gyro-resonant particle acceleration in the early afterglows of GRBs could account for the origin of UHECRs. It is worth noting that the evolution of jet's expansion can reduce the acceleration capacity of turbulence due to the dilution and adiabatic loss of the magnetic energy. In other words, the fluctuated magnetic field can energize cosmic-rays more efficiently without considering the evolution of jet's dynamics. For convenience, we list some parameters and their implications in the numerical calculation, as shown in~TABLE.\,\ref{table_1}. {}{Note that here we just show the spectra in four typical moments. We present the spectra at more moments of the evolution in Appendix.\,\ref{sec:Skeleton}.}

It is worth mentioning that, as an equally important component of the turbulent plasma, electrons might express non-thermal radiation processes in the early afterglows of GRBs. We believe that the electron acceleration in the frame of stochastic acceleration  has the value in itself. For example,
a study of SA of electron in the scenario of prompt emission of GRBs have been carried out to explain the origin of the Band function~\cite{Asano2015}. However, the focus of our present work is about UHECRs acceleration. The study of electron SA in the framework of our model will be studied in the near future.

\setlength{\arrayrulewidth}{0.2mm}
\setlength{\tabcolsep}{11pt}
\renewcommand{\arraystretch}{2}
\arrayrulecolor[HTML]{606060}

\begin{table*}[t]
\caption{The list of key parameters in numerical calculation.}\label{table_1}
\vspace{0.15cm}
\centering
\begin{threeparttable}[b]
\begin{tabular}{c c|c c}
  \hline\hline
  \multirow{1}*{\bf{Parameter}~~$($\bf{Units}$)$} & \bf{Definition} & \multicolumn{2}{c}{\bf{Value~~$($Cases$)$}}\\  \hline

  \multirow{1}*{$\xi$~~$\tnote{1}\,\,\,({\oslash})$}
  & $\textrm{dimensionless eddy scale}$ & \textrm{0.1}~$($I, II$)$ & \textrm{1}~$($III, IV$)$\\
    
  \multirow{1}*{$\nnn_{\textrm{ISM}}$~~$(\textrm{cm}^{-\textrm{3}})$}
  & $\textrm{number density of the homogeneous medium}$ & \textrm{0.01}~$($I, III$)$ & \textrm{1}~$($II, IV$)$ \\ 
  
  \multirow{1}*{$\BBB_{\textrm{0}}$~~$(\textrm{G})$}
  & $\textrm{initial magnetic field}$ & \textrm{1.84}~$($I, III$)$ & \textrm{18.4}~$($II, IV$)$\\
  
  \multirow{1}*{$\RRR_{\textrm{dec}}$~~$(\textrm{cm})$}
  & $\textrm{deceleration radius $($in comoving frame$)$}$  & $\textrm{5.6}\times\textrm{10}^{\textrm{17}}$~$($I, III$)$ & $\textrm{1.2}\times\textrm{10}^{\textrm{17}}$~$($II, IV$)$  \\ 
  
  \multirow{1}*{$\ttt_{\textrm{dec}}$~~$(\textrm{s})$}
  & $\textrm{deceleration time $($in comoving frame$)$}$  & $\textrm{7.8}\times\textrm{10}^{\textrm{4}}$~$($I, III$)$ & $\textrm{1.7}\times\textrm{10}^{\textrm{4}}$~$($II, IV$)$\\ 
  
  \multirow{1}*{$\alpha$~~$({\oslash})$}
  & $\textrm{the magnetic component of the total turbulent field}$ & \multicolumn{2}{c}{\textrm{0.25}~$($I\,--\,IV$)$}\\ 
    
  \multirow{1}*{$\mathcal{C}$~~$({\oslash})$}
  & $\textrm{Kolmogorov constant, appeared in Eq.~$($\ref{DKK}$)$}$ & \multicolumn{2}{c}{\textrm{1}~$($I\,--\,IV$)$}\\ 
  
  \multirow{1}*{$\tnote{2}\,\,\,\mathcal{E}_{\textrm{tot}}$~~$(\textrm{erg})$}
  & \textrm{total isotropic kinetic energy} & \multicolumn{2}{c}{$\textrm{10}^{\textrm{54}}$~$($I\,--\,IV$)$}\\
  
  \multirow{1}*{$\varepsilon_{\textrm{B}}$~~$(\oslash)$}
  & \textrm{magnetic field equipartition factor} & \multicolumn{2}{c}{$\textrm{0.025}$~$($I\,--\,IV$)$}\\ 
  
  \multirow{1}*{$\varepsilon_{\textrm{T}}$~~$(\oslash)$}
  & \textrm{equipartition factor of turbulent waves to $\EEE_{\textrm{tot}}$} & \multicolumn{2}{c}{$\textrm{0.1}$~$($I\,--\,IV$)$}\\
  
  \multirow{1}*{$\Gamma_{\textrm{0}}$~~$(${$\oslash$$)$}}
  & $\textrm{initial bulk Lorentz factor}$ & \multicolumn{2}{c}{\textrm{300}~$($I\,--\,IV$)$} \\ 
 
  \multirow{1}*{$\RRR_{\textrm{0}}$~~$(\textrm{cm})$}
  & $\textrm{initial radius of jet's evolution}$ & \multicolumn{2}{c}{$\textrm{5.4}\times\textrm{10}^{\textrm{14}}$~$($I\,--\,IV$)$}\\ 
  
  \multirow{1}*{\tnote{3}\,\,\,$\ttt_{\textrm{0}}$~~$(\textrm{s})$}
  & $\textrm{initial time of jet's evolution}$ & \multicolumn{2}{c}{\textrm{60}~$($I\,--\,IV$)$} \\ 
 
  \multirow{1}*{$\mathcal{D}_{\textrm{E}\textrm{E}}$~~$(\textrm{eV}^{\textrm{2}}$\,s$^{-\textrm{1}})$}  
  & \textrm{diffusion coefficient in energy space} &\multicolumn{2}{c} {\tnote{4}\,\,\,---~~~$($I\,--\,IV$)$}~~\\  
  
  \multirow{1}*{$\mathcal{D}_{\textrm{k}\textrm{k}}$~~$(\textrm{cm}^{-\textrm{2}}$\,s$^{-\textrm{1}})$}
  & \textrm{diffusion coefficient in wavenumber space} & \multicolumn{2}{c}{---~~~$($I\,--\,IV$)$}\\
  
  \multirow{1}*{$\Gamma_{\textrm{w}}$~~$(\textrm{s}^{-\textrm{1}})$}
  & $\textrm{damping rate of the cascading turbulent waves}$ & \multicolumn{2}{c}{---~~~$($I\,--\,IV$)$}\\ 
  
  \multirow{1}*{$\kkk_{\textrm{res,\,inj}}$~~$({\textrm{cm}}^{-\textrm{1}})$}
  & $\textrm{resonant injection wavenumber}$ & \multicolumn{2}{c}{---~~~$($I\,--\,IV$)$}\\
  
  \multirow{1}*{$\kkk_{\textrm{inj}}$~~$({\textrm{cm}}^{-\textrm{1}})$}
  & $\textrm{injection wavenumber}$ & \multicolumn{2}{c}{---~~~$($I\,--\,IV$)$}\\ 
  
  \multirow{1}*{$\WWW(\kkk)$~~$(\textrm{erg}\,\textrm{cm}^{-\textrm{2}})$}
  & $\textrm{total turbulence energy density per unit wavenumber}$ &  \multicolumn{2}{c}{---~~~$($I\,--\,IV$)$}\\
  
  \multirow{1}*{$\WWW_{\textrm{B}}(\kkk)$~~$(\textrm{erg}\,\textrm{cm}^{-\textrm{2}})$}
  & $\textrm{fast magnetosonic mode component of \WWW(\kkk)}$ & \multicolumn{2}{c}{---~~~$($I\,--\,IV$)$}\\
  
  \hline\hline
  
\end{tabular}
  \vspace{0.15cm}
   \begin{tablenotes}
     \item[1] Dimensionless physical parameter.
     \item[2] The initial energy of the burst measured by an observer is $\mathcal{E}_{\textrm{tot}}=\Gamma_{\textrm{0}}\textit{M}_{\textrm{ej}}c^{\textrm{2}}$.
     \item[3] At the phase of early afterglow, begin with $\textrm{0.1}$\,s after the burst in the frame of the central engine.
     \item[4] ``---'' means a set of data.
   \end{tablenotes}
\end{threeparttable}
\end{table*}

\section{Photodisintegration of UHECRs in the stochastic acceleration scenario}
\label{sec: Photodisintegration}
The information of the energy loss processes of nuclei can provide important clue to the mass composition of accelerated particles. An ultra-high energy nucleus with Lorentz factor $\Gamma_{\textrm{A}}$ traveling through an isotropic photon background with number density $\nnn_\gamma(\varepsilon_\gamma)\ddd\varepsilon_\gamma$ in the energy range $(\varepsilon_\gamma,\,\varepsilon_\gamma+\ddd\varepsilon_\gamma)$ suffers from loss of nucleons by the photodisintegration process, and the reaction rate is given by \cite{Stecker1968}$:$
  \begin{eqnarray}\label{photodisintegration}
  \ttt_{\textrm{dis}}^{\textrm{$-$1}} = \frac{\ccc}{{\textrm{2}}\Gamma_{\textrm{A}}^{\textrm{2}}}\int_{\varepsilon_{\textrm{th}}}^{\infty}\sigma_{\textrm{dis}}(\varepsilon_\gamma^{\prime})\varepsilon_\gamma^{\prime}\ddd\varepsilon_\gamma^{\prime}\int_{\varepsilon_\gamma^{\prime}/{\textrm{2}}\Gamma_{\textrm{A}}}^{\infty}\frac{\nnn_\gamma(\varepsilon_\gamma)}{\varepsilon_{\gamma}^{\textrm{2}}}\ddd\varepsilon_\gamma,~~
  \end{eqnarray}
  where $\ttt_{\textrm{dis}}$ represents the photodisintegration energy loss time, $\varepsilon_\gamma^{\prime}$ and $\varepsilon_{\gamma}$ are the photon energy in the nucleus rest frame and lab frame, respectively. The dominant channel of this process is called giant dipole resonance $($GDR$)$. The relevant threshold energy $\varepsilon_{\textrm{th}}=\textrm{10~MeV}$ and the cross section in the energy range $\varepsilon_\gamma^{\prime}\in({\varepsilon_{\textrm{th}}},\,\textrm{30~MeV})$ with loss of single nucleon can be roughly described in a Lorentzian form \cite{Puget1976}~as
  \begin{eqnarray}\label{cross_section1}
  \sigma_{\textrm{dis}}(\varepsilon_\gamma^{\prime})=\frac{\sigma_{\textrm{0}}{\varepsilon_\gamma^{\prime\textrm{2}}}\Delta_{\textrm{GDR}}^{\textrm{2}}}{(\varepsilon_{\textrm{0}}^{\textrm{2}}-{\varepsilon_\gamma^{\prime\textrm{2}}})^{\textrm{2}}+{\varepsilon_\gamma^{\prime\textrm{2}}}\Delta_{\textrm{GDR}}^{\textrm{2}}},
  \end{eqnarray}
  where $\sigma_{\textrm{\textrm{0}}}$ and $\Delta_{\textrm{GDR}}$ are the maximum value and width of the cross section with the peak energy $\varepsilon_{\textrm{0}}$. The numerical fitting values are $\sigma_{\textrm{0}}=\textrm{1.45}\,\textit{A}\times\textrm{10}^{-\textrm{27}}\,\textrm{cm}^{\textrm{2}}$, $\Delta_{\textrm{GDR}}=\textrm{8}\,\textrm{MeV}$, and $\varepsilon_{\textrm{0}}=\textrm{42.65}\,\textit{A}^{-\textrm{0.21}}\,\textrm{MeV}$ for $\textit{A}>\textrm{4}$ \cite{Karakula1993}. Eq.~$($\ref{cross_section1}$)$ is adequate for soft photon spectra. {}{Although a power-law function is more reasonable for the energy distribution of photon. However, for simplicity, we choose the delta function approximation $\sigma_{\textrm{dis}}(\varepsilon_\gamma^{\prime})\sim\sigma_{\textrm{0}}\Delta_{\textrm{GDR}}\delta(\varepsilon_\gamma^{\prime}-\varepsilon_{\textrm{0}})$ to estimate the reaction rate $($the results of estimation of these two methods are in the same order of magnitude, we can see that it will not affect our conclusion about the photodisintegration of heavier nuclei$)$}. 
  
  \begin{figure*}[ht!]
  \includegraphics[width=7.0in]{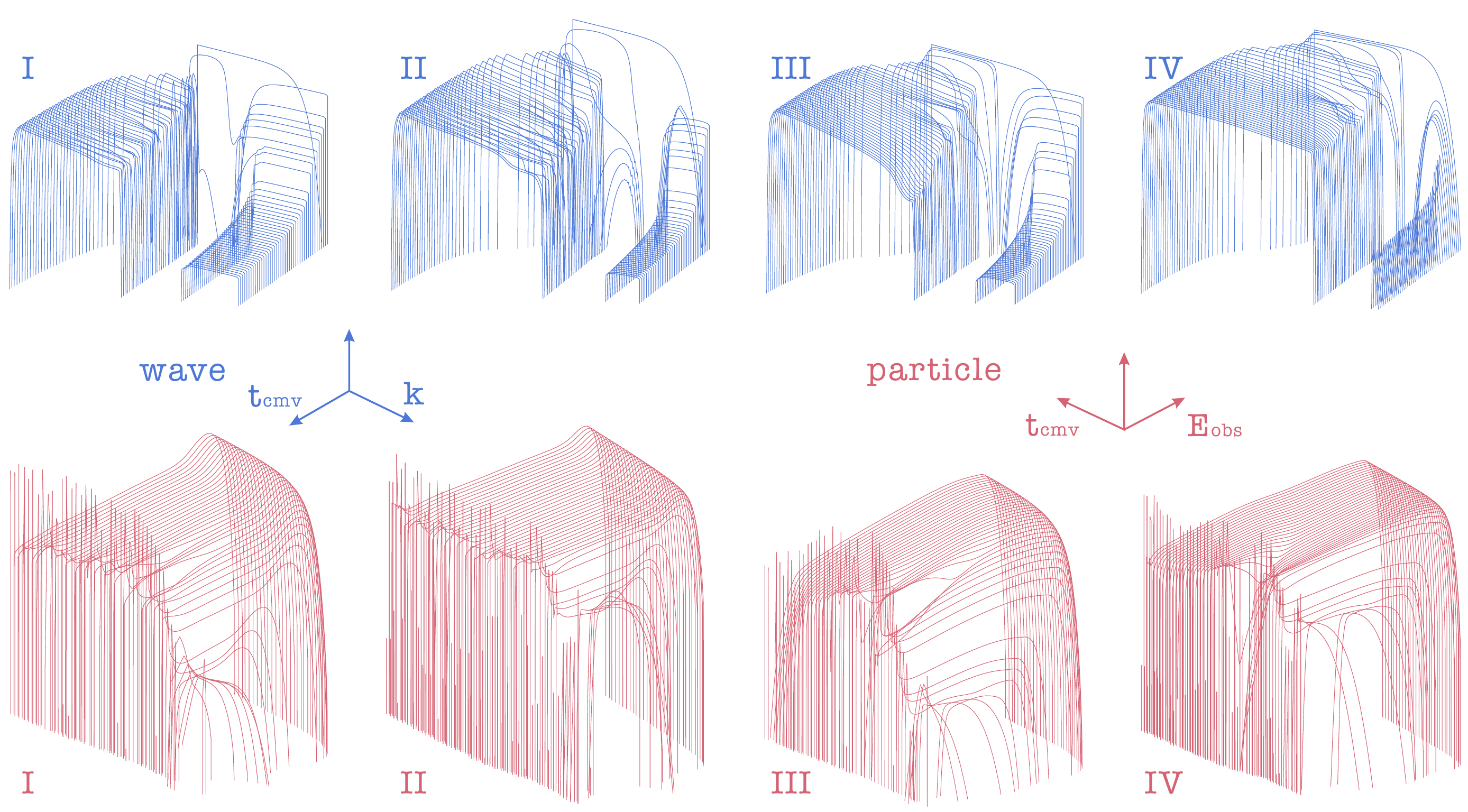}
  \caption{Upper panels: Four cases of turbulence spectral energy density spectra evolution. Lower panels: Four cases of UHECR protons spectra evolution. The time vector of the triads represent the evolution direction.}
  \label{fig:App1}
\end{figure*}
  
  The accelerated ultra-high energy nuclei with {energies} $\varepsilon_{\textrm{obs}}$ above $\textrm{10}^{\textrm{19}}\,\textrm{eV}$ prefer to interact with these X-ray photons if the shock's Lorentz factor $\Gamma\in(\textrm{10}^{\textrm{2}},\,\textrm{10}^{\textrm{3}})$, and the Lorentz factor of an ultra-high energy nucleus $\Gamma_{\textrm{A}}=\varepsilon_{\textrm{obs}}/(\AAA\Gamma\mmm_{\textrm{p}}\ccc^{\textrm{2}})$ in the observer's frame. Assuming the spectrum of the early X-ray afterglow follows the fast-cooling behavior with $\textit{F}_{\nu}\sim{\nu}^{\textrm{$-$1}}$ \cite{Peter2006}, then we can get the photodisintegration rate of a nucleus moving with $\Gamma_{\textrm{A}}$ \cite{Wang2008}$:$
  \begin{eqnarray}\label{dis_rate}
  \ttt_{\textrm{dis}}^{\textrm{$-$1}} = \frac{\textrm{4}}{\textrm{3}}\ccc\sigma_{\textrm{0}}\frac{\Delta_{\textrm{GDR}}}{\varepsilon_{\textrm{0}}^{\prime}}\frac{\Gamma_{\textrm{A}}\textit{U}_{\textrm{X}}}{\kappa\varepsilon_{\textrm{0}}^{\prime}},
  \end{eqnarray}
  where $\textit{U}_{\textrm{X}}=\kappa\nnn_{\textrm{b}}(\varepsilon_{\textrm{b}})\varepsilon_{\textrm{b}}^{\textrm{2}}$ is the comoving-frame energy density of X-ray afterglow photons and $\varepsilon_{\textrm{b}}$ the break energy, $\kappa=\textrm{ln}(\varepsilon_{\textrm{BAT},+}/\varepsilon_{\textrm{BAT},-})\simeq\textrm{3}$, with $\varepsilon_{\textrm{BAT},+}$ and $\varepsilon_{\textrm{BAT},-}$ being the upper and lower end of \textit{Swift}-BAT energy threshold. In the early phase of the external shock for a GRB with bright X-ray afterglow emission, such as GRB\,\textrm{190114C} \cite{MAGIC2019}, the average luminosity of the relevant X-ray afterglow observed by \textit{Swift}-BAT is about $\textit{L}_{\textrm{X}}=\textrm{4}\pi \RRR_{\textrm{ex}}^{\textrm{2}}\Gamma^{\textrm{2}}\ccc\textit{U}_{\textrm{X}}\simeq\textrm{10}^{\textrm{48.5}}\,\textrm{erg}\,\textrm{s}^{\textrm{$-$1}}$ during the initial $\sim\textrm{68}-\textrm{110}$\,s, where $\RRR_{\textrm{ex}}$ is the radius of the external shock at the final stage of the free expansion phase of the turbulent ejecta. For ultra-high energy nuclei, the effective optical depth $\tau=\ttt_{\textrm{dyn}}/\ttt_{\textrm{dis}}$ for photodisintegration with four different cases that mentioned above at $\ttt_{\textrm{obs}}\simeq\textrm{80}~\textrm{s}$ are given by
\begin{eqnarray}\label{dis1}
  \begin{cases}
{}{\textrm{6.5}\times\textrm{10}^{-\textrm{5}}}~\textit{L}_{\textrm{X},\textrm{48.5}}\textit{E}_{\textrm{obs},{}{\textrm{18}}}\RRR_{\textrm{ex},\textrm{17.5}}^{-\textrm{1}}\Gamma_{\textrm{249}}^{-\textrm{4}}\big({\AAA}/{\textrm{56}}\big)^{\textrm{0.42}}
&\textbf{I}\\

{}{\textrm{6.5}\times\textrm{10}^{-\textrm{4}}}~\textit{L}_{\textrm{X},\textrm{48.5}}\textit{E}_{\textrm{obs},{}{\textrm{19}}}\RRR_{\textrm{ex},\textrm{17.5}}^{-\textrm{1}}\Gamma_{\textrm{249}}^{-\textrm{4}}\big({\AAA}/{\textrm{56}}\big)^{\textrm{0.42}}
&\textbf{III}\\

{}{\textrm{9.1}\times\textrm{10}^{-\textrm{3}}}~\textit{L}_{\textrm{X},\textrm{48.5}}\textit{E}_{\textrm{obs},{}{\textrm{18.5}}}\RRR_{\textrm{ex},\textrm{17.1}}^{-\textrm{1}}\Gamma_{\textrm{123}}^{-\textrm{4}}\big({\AAA}/{\textrm{56}}\big)^{\textrm{0.42}}
&\textbf{II}\\

{}{\textrm{9.1}\times\textrm{10}^{-\textrm{2}}}~\textit{L}_{\textrm{X},\textrm{48.5}}\textit{E}_{\textrm{obs},{}{\textrm{19.5}}}\RRR_{\textrm{ex},\textrm{17.1}}^{-\textrm{1}}\Gamma_{\textrm{123}}^{-\textrm{4}}\big({\AAA}/{\textrm{56}}\big)^{\textrm{0.42}}
&\textbf{IV}
\end{cases}.~~~~~~
  \end{eqnarray}
From the above results, we conclude that for all the four different cases, the ultra-high energy nuclei $($iron$)$ can easily survive photodisintegration. 
From Hillas criterion, we know that the maximum energy of UHECR is $\textit{E}_{\textrm{max}}\propto\textit{AeBl}$, where $\textit{l}$ is the scale of acceleration region. As long as ultra-high energy nuclei survive photodisintegration, heavier nuclei can achieve higher maximum energy.

\section{Conclusions}
\label{sec: Conclusions}
In this paper, we take into account the concurrence of GRB jet's dynamics and the kinetic descriptions of wave-particle interactions including  SA  process of particles and the damping of MHD fast-mode waves. Protons can be accelerated to ultra-high energy by turbulent waves through wave-particle gyro-resonant interactions.
 
Including the evolution of jet's dynamics can reduce the energy density of the turbulent magnetic fields, and subsequently weaken the capacity of the acceleration of the SA mechanism. Since energies of accelerated particles originate from the magnetic turbulence, taking into account the feedback (i.e., damping) of particle acceleration on the turbulence spectrum leads to a weaker magnetic field compared to that predicted in the standard afterglow dynamic model, given that the magnetic energy is consumed by particles. It also results in a particle spectrum softer than that predicted in the test-particle limit. Considering the fast mode of magnetosonic wave as the dominant particle scatterer and assuming ISM for the circumburst environment, we found that protons can nevertheless be accelerated up to $\textrm{10}^{\textrm{19}}\,$eV with a spectrum $\textrm{d}\NNN/\textrm{d}\EEE \propto \EEE^{-\textrm{1}}$ for some favorable choices of system's parameters. We also found that a pile-up bump may occur in the spectrum ahead of the cutoff, if the injection eddy scale is small, leading to a very hard particle spectrum with $\textrm{d}\NNN/\textrm{d}\EEE \propto \EEE^{\textrm{\,0}}$. On the other hand, the maximum energy (or cutoff energy) of the accelerated protons is reduced because the maximum achievable energy in the acceleration is limited by the eddy scale.

An analytic estimate shows that the ultra-high energy nuclei can easily survive photodisintegration in the early afterglows of GRBs, which imply intermediate-mass or heavy nuclei can achieve $\textrm{10}^{\textrm{20}}\,$eV in our model if they are loaded in GRB jets. Compared to the traditional acceleration model by relativistic shocks, our model not only alleviates the energy budget problem, but also provide a mechanism to 
generate the hard injection spectrum as required by explaining the measured UHECRs spectra above the ankle and the chemical composition of UHECR as measured by the Pierre Auger Observatory.

\begin{figure*}[ht!]
  \includegraphics[width=6.8in]{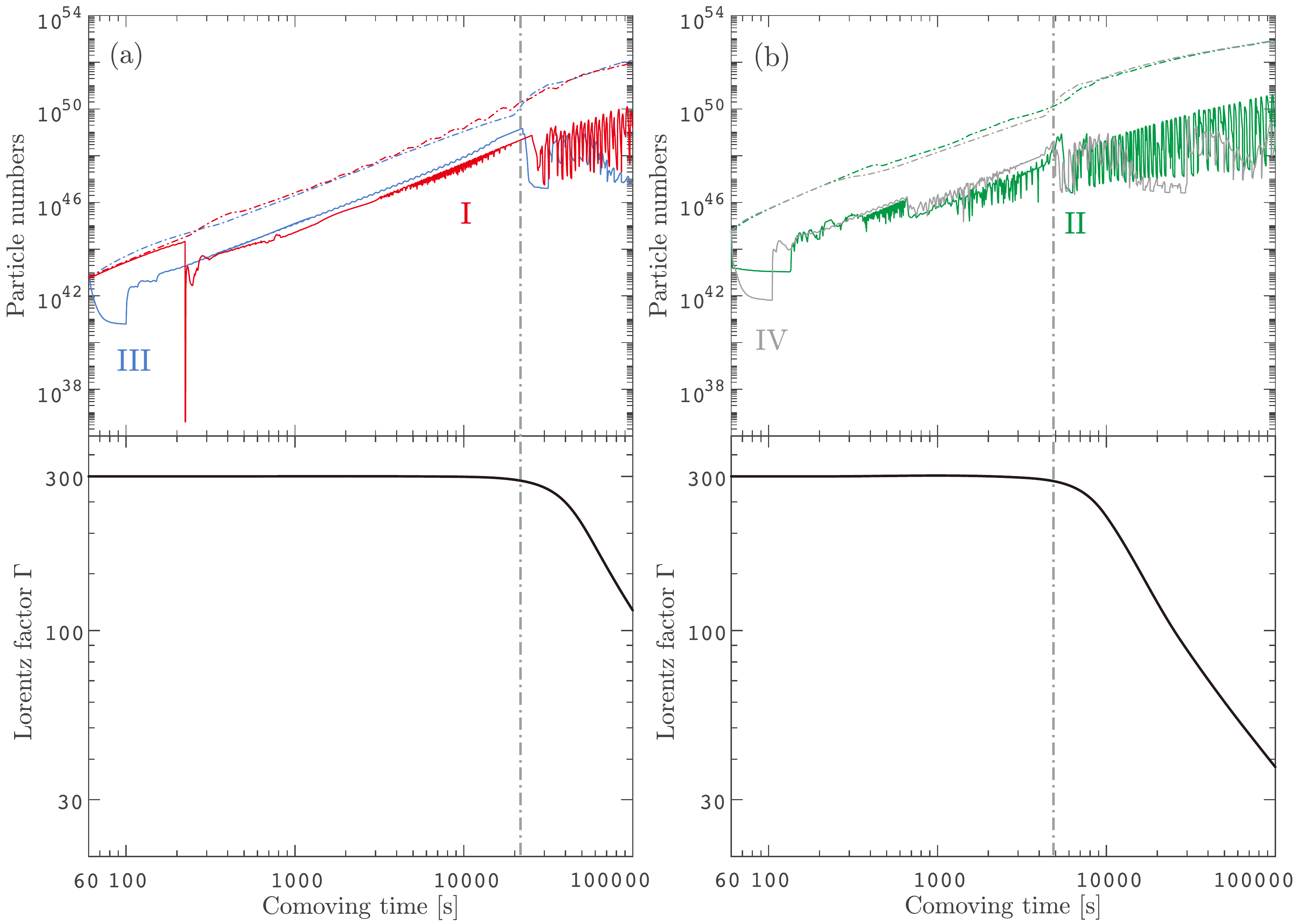}
  \caption{Upper panels: The number of protons evolution in four different cases. The red lines illustrate case \textbf{I}, the blue lines illustrate case \textbf{III}, the green lines illustrate case \textbf{II}, and the gray lines illustrate case \textbf{IV}. Dash-dotted lines do not take the spatial escape effect into account, while the solid lines do. Lower panels: The evolution of the bulk Lorentz factor $\Gamma$ of the shock wave in the jet's comoving frame. The left panel shows the cases \textbf{I} and \textbf{III}, while the right panel shows the cases \textbf{II} and \textbf{IV}. The vertical dash-dotted lines represent the moment when $\Gamma$ starts to decrease significantly.}
  \label{fig:App2}
\end{figure*}

\begin{acknowledgments}

We thank the anonymous referee for the constructive report that improved the quality of this paper. We also acknowledge helpful discussions with Peter M{\'e}sz{\'a}ros, Katsuaki Asano, Huirong Yan, Joshi Jagdish and Jun Kakuwa. 
This work is supported by the National Key R \& D program of China under the Grant No. 2018YFA0404203 and the NSFC Grants Nos. 11625312, 11851304, U2031105.

\end{acknowledgments}

\begin{appendix}
\section{Skeleton plots of Fig.\,\ref{fig:Fig1}, Fig.\,\ref{fig:Fig2} and Fig.\,\ref{fig:Fig4}}\label{sec:Skeleton}
In order to illustrate the damping of turbulent waves not only occurs around larger wavenumbers, but also occurs around smaller wavenumbers, we plot more moments for four different cases of the UHECR protons spectra and the turbulence spectra, as shown in Fig.\,\ref{fig:App1}. We can see that at very early stage of the wave-particle system's evolution in cases \textbf{I} and \textbf{II}, the wave damped by protons significantly around $\kkk_{\textrm{inj}}$, it means almost all of the turbulence energy is extracted by protons via gyro-resonant interactions. That is also the reason why the magnetic field energy density in Fig.\,\ref{fig:Fig3} drops such quickly than that in the other two cases.

From Fig.\,\ref{fig:App1}, we notice that the quasi-periodic fluctuation behaviour around injection energy on the proton spectra and around the resonant injection wavenumber $($sometimes at high wavenumbers and sometimes at low wavenumbers --- even around the injection of waves $\kkk_{\textrm{inj}}${}$)$ on the turbulence spectra. This behavior is caused by the resonant wave-particle interactions.

\section{The number of protons evolution}\label{sec:number}
The number of particles evolution in jet's comoving frame under four different cases. The fluctuations on the curves are induced by the joint effects of the wave-particle gyro-resonant interactions, adiabatic cooling of turbulent magnetic fields and particles escape, as shown in Fig.\,\ref{fig:App2}. In the absence of particle escape, the number of particles continuously increase until the end of the evolution. However, in the case of particle escape $($cases \textbf{II}, \textbf{III} and \textbf{IV}$)$, the energized particles which extract energy from turbulent waves will escape from the acceleration region, causing the number of these particles drops until about a hundred seconds in the comoving frame. The reduction in particle number also reduces the damping rate. After then for a while, the newly injected magnetic energy gradually increase to a certain amount which can keep dynamic quasi-equilibrium with the adiabatic cooling of themselves and the damping of waves by particles until the bulk Lorentz factor $\Gamma$ of the shock begins to drop significantly, as shown in the lower panels of Fig.\,\ref{fig:App2}. Due to the high sensitivity to the variation of the value of $\Gamma$, the evolution of the non-linear coupled FP equations will going to enter the second dynamic equilibrium process. The multiplicity of the fluctuation of the number of particles evolution originates from the feature of the logarithmic coordinate and the decline of $\Gamma$. The interpretation of the peculiarity of case \textbf{I} can be found in the bottom panels of Fig.\,\ref{fig:Fig3} and Appendix.\,\ref{sec:vs}.

\section{The injection wavenumber $\kkk_{\textrm{inj}}$ vs. the resonant injection wavenumber $\kkk_{\textrm{res,\,inj}}$}\label{sec:vs}

When $\kkk_{\textrm{inj}}>\kkk_{\textrm{res,\,inj}}$, the following condition should be met
\begin{eqnarray}\label{Bksi}
\BBB\xi>\eta\equiv\frac{\textrm{2}\pi\Gamma\EEE_{\textrm{inj}}}{e\RRR}.
\end{eqnarray}
From Fig.\,\ref{fig:Fig3}, we can see that the condition is well satisfied in cases \textbf{II}, \textbf{III} and \textbf{IV}. However, in case \textbf{I}, $\kkk_{\textrm{res,\,inj}}=\kkk_{\textrm{inj}}$ at very early stage of the evolution $(${}$\ttt_{\textrm{0}}=\textrm{60}$\,s$)$ for the first time in the jet's comoving frame. The value of $\eta$ remains $\textrm{0.0033}$ at the early stage of the evolution. The initial value of $\BBB\xi=\BBB_\textrm{0}\xi\simeq\textrm{0.1844}$ is larger than the value of $\eta$ at the beginning of the evolution until $\textrm{60.6995}\,$s. The damping of turbulent waves occurs around the injection wavenumber until $\kkk_{\textrm{res,\,inj}}=\kkk_{\textrm{inj}}$ again around $\textrm{223.5}\,$s in the comoving frame, as shown in the bottom panels of Fig.\,\ref{fig:Fig3}. We know that the initial magnetic field $\BBB_{\textrm{0}}\propto\nnn_{\textrm{ISM}}^{\textrm{1/2}}$ with $\nnn_{\textrm{ISM}}=\textrm{0.01}\,{\textrm{cm}}^{-\textrm{3}}$ and $\xi=\textrm{0.1}$ in case \textbf{I}, as damping occurs at the injection scale, the magnetic fields drop quickly due to the adiabatic cooling of themselves and the damping of waves by particles, the value of $\BBB\xi$ is more likely to turn smaller than the value of $\eta$ than other three cases. With the further injection of fast magnetosonic waves, the decline of resonant wavenumber is very slow until $\kkk_{\textrm{res,\,inj}}=\kkk_{\textrm{inj}}$ again at $\ttt_{\textrm{cmv}}=\textrm{223.5}\,$s. Actually, there is no turbulent waves to energize particles when $\kkk_{\textrm{res,\,inj}}<\kkk_{\textrm{inj}}$. Therefore, during the period from $\ttt_{\textrm{cmv}}=\textrm{60.6995}\,$s to $\ttt_{\textrm{cmv}}=\textrm{223.5}\,$s, the injected particles are not accelerated to higher energy, consequently, the number of particles remains unchanged, as shown in case \textbf{I} of Fig.\,\ref{fig:App2} in Appendix.\,\ref{sec:number}. 

After the ``step'' transition $($from $\kkk_{\textrm{res,\,inj}}>\kkk_{\textrm{inj}}$ to $\kkk_{\textrm{res,\,inj}}<\kkk_{\textrm{inj}}$ again$)$ in case \textbf{I}, the newly injected magnetic energy accumulates very soon, resulting in a tiny bump at the moment. In the meantime, the accumulated particles can gain energy from turbulent waves via gyro-resonant interactions again. Thus the escape effect of particles is significant within a very short period of time, as shown in Appendix.\,\ref{sec:number}. This is also the reason for the nontrivial behaviors of the evolution of magnetic field energy density and turbulence energy and cosmic-ray energy of case \textbf{I} in Fig.\,\ref{fig:Fig3}.

\end{appendix}

\end{document}